\documentclass[twocolumn,amsmath,amssymb,aps,preprintnumbers,prd]{revtex4-1}
\usepackage{graphicx} % Allows for eps images
\usepackage{color}
\usepackage{slashed} 
\usepackage{hyperref}
\allowdisplaybreaks
\def\beq{\begin{equation}}
\def\eeq{\end{equation}}
\def\beqn{\begin{eqnarray}} 
\def\eeqn{\end{eqnarray}}
\def\nn{\nonumber}

\def\uv{\mathrm{UV}}

\newcommand{\salamanca}{Departamento de F\'isica Fundamental e IUFFyM, Universidad de Salamanca, Plaza de la Merced S/N, 37008 Salamanca, Spain.}
\newcommand{\ueuropea}{GFIMA - Escuela de Ciencias, Ingenier\'ia y Diseño, Universidad Europea de Valencia, Paseo de la Alameda 7, 46010 Valencia, Spain.}

\begin{document}
\title{Towards multiloop local renormalization within Causal Loop-Tree Duality}

\author{Jos\'e R\'ios-S\'anchez$^{a}$, German Sborlini$^{a,b}$}
\affiliation{$^{a}$ \salamanca\\
$^{b}$ \ueuropea}

\begin{abstract}
Renormalization is a well-known technique to get rid of ultraviolet (UV) singularities. When relying on Dimensional Regularization (DREG), these become manifest as $\epsilon$-poles, allowing to define counter-terms with useful recursive properties. However, this procedure requires to work at \emph{integral-level} and poses difficulties to achieve a smooth combination with semi-numerical approaches. This article is devoted to the development of an \emph{integrand-level} renormalization formalism, better suited for semi or fully numerical calculations. Starting from the Loop-Tree Duality (LTD), we keep the causal representations of the integrands of multiloop Feynman diagrams and explore their UV behaviour. Then, we propose a strategy that allows to build local counter-terms, capable of rendering the expressions integrable in the high-energy limit and in four space-time dimensions. Our procedure was tested on diagrams up to three-loops, and we found a remarkably smooth cancellation of divergences. The results of this work constitute a powerful step towards a fully local renormalization framework in QFT.
\end{abstract}

\setcounter{page}{1}
\maketitle

%%%%%%%%%%%%%%%%%%%%%%%%%%%%%%%%%%%%%%%%%%%%%%%%%%%%%%%%%%%%%%%%%%%%%%%%%%%%%%%%%%%%%%
%%%%%%%%%%%%%%%%%%%%%%%%%%%%%%%%%%%%%%%%%%%%%%%%%%%%%%%%%%%%%%%%%%%%%%%%%%%%%%%%%%%%%%
\section{Introduction and motivation}
\label{sec:Introduction}
In the quest for precision in high-energy physics, the need to compute higher perturbative orders has become essential. Current and future high-energy colliders will collect an enormous amount of data, which will lead to very precise measurements. From the theory side, this translates into a huge challenge to reduce the uncertainties in the predictions extracted from Quantum Field Theories (QFT) and pushes the available computational frameworks to their limits.

In spite of the recent developments of several new methodologies to tackle higher-order calculations \cite{Heinrich:2020ybq}, there are some bottlenecks reluctant to be solved. One of them is related to the presence of divergences and their non-local cancellation. It is well-known that generic QFTs possess singularities in the high-energy limit, the so-called ultraviolet (UV) divergences. For the particular case of gauge theories, it is also known \cite{tHooft:1972tcz,Collins:1984xc} that these divergences can be always absorbed or hidden within free parameters of the theory, through the renormalization procedure. On the other hand, QFT
with massless particles also have low-energy divergences, known as infrared (IR) singularities. As in the UV case, there are several well-established methods to tackle them, which mostly consist in adding and removing suitable counter-terms \footnote{We refer the interested reader to Refs. \cite{Gnendiger:2017pys,TorresBobadilla:2020ekr} for updated, although not fully complete, reviews on the topic.}. 

Both for IR and UV singularities, it is required to introduce a regularization prescription to make manifest and handle the divergences. Due to its nice properties, Dimensional Regularization (DREG) \cite{Bollini:1972ui,Ashmore:1972uj,tHooft:1972tcz,Cicuta:1972jf} has become a rather standard approach. When working with $D=4-2\epsilon$ space-time dimensions, the singularities manifest as poles in $\epsilon$. In the IR sector, the $\epsilon$ poles cancel when we consider IR-safe observables and we put together all the degenerated configurations that include extra real-radiation and virtual particles \cite{Kinoshita:1962ur}. Within the traditional subtraction approaches, this cancellation occurs \emph{after} integration of the real-radiation and loop amplitudes, including some suitable counter-terms. Analogously, the UV divergences are removed through renormalization counter-terms, which include $\epsilon$-poles that exactly match those present in the virtual amplitudes. At this point, we have to emphasize that UV divergences are only originated inside the loops, due to the fact that the energy of the virtual states is unconstrained.

The computational framework described before has been successfully applied to several relevant processes in collider physics, up to next-to-next-to-leading order (NNLO), which has become the \emph{new standard} in precision. Very recently, next-to-next-to-next-to-leading order (NNNLO) results appeared \footnote{For instance, see Ref. \cite{Baglio:2022wzu} and references therein.}, although the effort required to achieve them is increasing enormously when adding extra loops or legs. Among the different bottlenecks, there are severe difficulties to analytically calculate multi-loop multi-leg Feynman integrals within DREG, since the presence of IR and UV singularities prevents a straightforward numerical implementation. Besides, IR divergences within loops avoid a direct numerical cancellation with those present in the real-emission contributions, forcing to use non-local or semi-analytical techniques to render the expressions IR-finite.

With this panorama in mind, novel techniques to explore an efficient point-by-point or local cancellation of singularities \emph{before} integration are required \cite{Anastasiou:2018rib,Anastasiou:2020sdt,Anastasiou:2022eym,Sterman:2023xdj}. In this direction, the Loop-Tree Duality (LTD) \cite{Catani:2008xa,Bierenbaum:2010cy,Bierenbaum:2012th,Buchta:2014dfa,Buchta:2015wna,Plenter:2019jyj,Plenter:2020lop,Plenter:2022zxk,Runkel:2019zbm,Aguilera-Verdugo:2021nrn} allows to open loop amplitudes into sums of tree-level-like objects integrated over a phase-space, closely resembling the real-radiation contribution. Using this formalism, the IR-singular structure of the loops is expressed in terms of phase-space integrals \cite{Buchta:2014dfa}, suggesting a clear connection to the real-radiation contribution. In fact, this constitutes the basis of the Four Dimensional Unsubtraction (FDU) framework \cite{Hernandez-Pinto:2015ysa,Sborlini:2016gbr,Sborlini:2016hat,Prisco:2020kyb}, where the \emph{open loops} or dual amplitudes are combined with the real-radiation by means of suitable momentum mappings. As a result, FDU provides integrand-level expressions that are explicitly free of IR singularities, which are cancelled locally (i.e. point-by-point) \emph{before} integration, rendering the expression not only finite but also (and most importantly) integrable.

In order to achieve a local cancellation of UV divergences, there are some recent methodologies in the market \cite{Pittau:2012zd,Donati:2013voa,Pittau:2014tva,Page:2015zca,Gnendiger:2017pys,TorresBobadilla:2020ekr}, although most of them are not optimized for local IR counter-terms. Within the FDU framework, the treatment of UV singularities is slightly different, but the objective is the same: develop local renormalization counter-terms. Even if the well-known Bogoliubov-Parasiuk-Hepp-Zimmermann (BHPZ) renormalization programme \cite{Bogoliubov:1957gp,Hepp:1966eg,Zimmermann:1967XXX, Zimmermann:1969jj,Lowenstein:1974qt,Gomes:1974cr,Lowenstein:1975rg,Lowenstein:1975ps,Lowenstein:1975ug,Blaschke:2013cba} provides a systematic way of computing local UV subtraction counter-terms, it was mainly thought for renormalization in Minkowski space and requires some adjustments to be incorporated within the FDU framework. Our procedure is inspired in the expansion around UV-propagators \cite{Becker:2010ng,Driencourt-Mangin:2019aix,Driencourt-Mangin:2019sfl,Driencourt-Mangin:2019yhu}, combined with the application of the LTD and a suitable matching procedure to recover results in DREG-defined schemes (such as $\overline{\rm MS}$). In this way, FDU has proven able to achieve a fully local cancellation of both IR and UV divergences up to next-to-leading order (NLO). Still, local UV renormalization in FDU beyond NLO is an open problem. A strategy to deal with UV singularities up to two-loop was developed in Refs. \cite{Driencourt-Mangin:2019aix,Driencourt-Mangin:2019yhu}, allowing to numerically reproduce some well-known $\overline{\rm MS}$ results.

In this article, we explore a LTD-based strategy that makes use of the so-called causal representation \cite{Tomboulis:2017rvd,Aguilera-Verdugo:2019kbz,Verdugo:2020kzh,Aguilera-Verdugo:2020kzc,Ramirez-Uribe:2020hes,Aguilera-Verdugo:2020nrp,snowmass2020,Runkel:2019yrs,Capatti:2019ypt,Capatti:2020ytd} to unveil local UV counter-terms for multi-loop Feynman integrals. The methodology is inspired, again, in the expansion around UV-propagators but acting directly on the Euclidean space of the causal dual integrands (i.e. the integrands of the Feynman amplitudes after the application of the manifestly causal LTD). We used the fact that the causal dual representations has an analogous structure to the real-radiation phase-space, allowing to split the integrand in such a way that we can isolate the UV divergent contributions more efficiently.

The outline of this article is the following. In Sec. \ref{sec:CausalLTD}, we briefly review the basis of manifestly causal LTD and its mathematical properties. Then, we discuss the generation of local UV counter-terms in Sec. \ref{sec:LocalReno}. We start reviewing in Sec. \ref{ssec:UVFelix} the strategy presented in Refs. \cite{Driencourt-Mangin:2019aix,Driencourt-Mangin:2019sfl}, and we explain the causal inspired-approach in Sec. \ref{ssec:CausalLTD}. After presenting the general formalism, we explore useful simplifications to keep the causal structure in the local counter-terms in Sec. \ref{ssec:onshellsimple}. In Sec. \ref{sec:Examples}, we apply our renormalization methodology to two and three-loop representative diagrams, showing a smooth convergence in the UV region. After discussing the results within our method, we provide a comparison with BPHZ renormalization program in Sec. \ref{sec:BPHZ}, highlighting the similarities and differences. Also, we include in App. \ref{app:A} an analysis about the momentum expansions leading to the UV counter-terms in both approaches. Finally, we present the conclusions and future research directions in Sec. \ref{sec:Conclusions}.

%20231217: OK!!
%20240117: OK, mejorado!!
%20240427: INCORPORADAS CORRECCIONES

%%%%%%%%%%%%%%%%%%%%%%%%%%%%%%%%%%%%%%%%%%%%%%%%%%%%%%%%%%%%%%%%%%%%%%%%%%%%%%%%%%%%%%
\section{Causal Loop-Tree Duality}
\label{sec:CausalLTD}
The motivation behind the Loop-Tree Duality (LTD) is intuitive: open loop diagrams into a collection of tree-level like objects. To achieve this purpose, we make use of Cauchy's residue theorem to remove one degree-of-freedom per loop. In particular, we choose to integrate out the energy component of each loop momenta, which transforms the integration domain from a Minkowski to an Euclidean space. This point is very important in order to reach an integrand-level combination of the different ingredients involved in higher-order cross-section computations (namely, the real-radiation and the virtual corrections). Furthermore, integrating out the energy component implies putting on-shell certain internal lines of the diagrams.

In its original form, LTD decomposes any one and two-loop scattering amplitude in any QFT into a sum of trees where some subsets of propagators were replaced by the so-called \emph{dual propagators} \cite{Catani:2008xa,Rodrigo:2008fp,Bierenbaum:2010cy,Bierenbaum:2012th}. The dual propagators, denoted $G_D(q_i;q_j)$, have a momentum dependent prescription that allows to capture all the information contained in the multiple cuts originated from the Feynman Tree Theorem (FTT) \cite{Feynman:1963ax}. The integrand obtained after the application of the LTD is called \emph{dual integrand} or \emph{dual representation} of the associated Feynman integral or multi-loop scattering amplitude. 

Recently, it was found that the iterated application of the Cauchy residue theorem directly leads to the dual representation. To illustrate this, let us consider a generic $L$-loop scattering amplitude with $N$ external particles,
\beqn
\nn {\cal A}_N^{(L)} &=& \int_{\ell_1 \ldots \ell_L} \, \sum_j \, {\cal N}_j \times G_F(1, \ldots, L, \ldots, n) 
\\ && \approx \int_{\ell_1 \ldots \ell_L} d{\cal A}_N^{(L)} (1,\ldots,n)\, ,
\label{eq:GenericAmplitude}
\eeqn
where ${\cal N}$ corresponds to the numerator (depending on the loop and external momenta, $\{\ell_i\}_{i=1,\ldots,L}$ and $\{p_i\}_{i=1,\ldots,N}$, respectively), and $G_F(1, \ldots,L,\ldots, n)$ denotes a product of Feynman propagators associated to the underlying topology. $\{1,\ldots,n\}$ denotes sets of internal momenta that depend on specific combinations of loop momenta. For the sake of simplicity, we can assume $1$ depends only on $\ell_1$, $2$ only on $\ell_2$ and so on, whilst the remaining are non-trivial linear combinations of $\{\ell_i\}_{i=1,\ldots,L}$. Regarding the integration measure, we have:
\beq
\int_{\ell_1 \ldots \ell_L} = (-i \, \mu^{4-d})^L \, \int \frac{d^d\ell_1}{(2\pi)^d} \ldots \frac{d^d\ell_L}{(2\pi)^d} \, ,
\eeq
which is valid for an arbitrary number of space-time dimensions $d$, and $\mu$ denotes an energy scale associated to the regularization procedure (DREG in this case). 

We start by taking the residue over the energy component of $\ell_1$, which is equivalent to put on-shell, one-by-one, the propagators of $1$, i.e. 
\beq
{\cal A}^{(L)}_D(1;2,\ldots,n) = \sum_{i_1 \in 1} {\rm Res}\left(d{\cal A}_N^{(L)}(1,\ldots,n);q_{i_1,0}^{(+)}\right) \, ,
\label{eq:AmplitudNLGeneral}
\eeq
where $q_{j,0}^{(+)}=\sqrt{\vec{q_j}^2+m_j^2-i0}$ is the positive on-shell energy of the $j$-th line. Then, we iterate the procedure and we get
\beqn
\nn && {\cal A}^{(L)}_D(1,2,\ldots,k;k+1,\ldots,n) = 
\\ && \sum_{i_k \in k} {\rm Res}\left({\cal A}^{(L)}_D(1,2,\ldots,k-1;k,\ldots,n);q_{i_k,0}^{(+)}\right) \, ,
\eeqn
in the $k$-th step. The dual representation for an $L$-loop amplitude is obtained by computing the $L$-th iterated residue, and the original scattering amplitude is then written
\beqn
{\cal A}_N^{(L)} &=& \int_{\vec{\ell}_1 \ldots \vec{\ell}_L} 
\, {\cal A}^{(L)}_D(1,2,\ldots,k;k+1,\ldots,n) \, ,
\label{eq:DualRepresentation}
\eeqn
with the integration being performed over an Euclidean space (defined by the tensorial product of the space components of the loop momenta) and
\beq
\int_{\vec{\ell}_i} = \mu^{4-d} \int \frac{d^{d-1}\vec{\ell_i}}{(2\pi)^{d-1}}\, .
\label{eq:IntegrationMeasure1}
\eeq
It is worth appreciating that several contributions vanished when iterating the residues; these are associated to the so-called \emph{displaced poles}, which correspond to non-physical configurations and allow us to define the concept of \emph{nested residues} \cite{Aguilera-Verdugo:2021nrn}.

Remarkably, further simplifications take place when all the dual contributions (i.e. the terms inside ${\cal A}_D^{(L)}$) are explicitly added together: only those terms compatible with causality survive \cite{Aguilera-Verdugo:2020kzc}. In fact, Eq. (\ref{eq:DualRepresentation}) can be re-casted as
\beqn 
\nonumber {\cal A}_N^{(L)} &=& \sum_{\sigma \in \Sigma} \int_{\vec{\ell}_1 \ldots \vec{\ell}_L} (-1)^k\, \frac{{\cal N}_{\sigma}(\{q_{r,0}^{(+)}\},\{p_{j,0}\})}{x_{L+k}} \, 
\\ &\times& \prod_{i=1}^k \frac{1}{\lambda_{\sigma(i)}} \, + \, (\sigma \longleftrightarrow \bar{\sigma}) \, ,
\label{eq:DualCausalRepresentation}
\eeqn
where we introduced the following definitions:
\begin{itemize}
\item The integration measure, analogous to the real-emission phase space:
\beq
x_{L+k} = \prod_{j=1}^{L+k} 2 q^{(+)}_{j,0} \, .
\label{eq:Prefactor}
\eeq
\item The order of the topology associated to the Feynman graph $k=V-1$ (with $V$ the number of vertices).
\item The causal propagators $1/\lambda_i$, with $\{\lambda_i\}$ are associated to physical threshold singularities of the diagram.
\end{itemize}
Also, $\sigma$ represents a combination of $k$ \emph{entangled} physical thresholds from the set $\Sigma$ of all the compatible thresholds. It turns out that this formulation only contains terms that become singular when going through physical thresholds, manifestly exploiting the underlying causality of QFT and motivating its name: \emph{causal dual representation} \cite{Verdugo:2020kzh,TorresBobadilla:2021ivx,Sborlini:2021owe}. It is important to highlight that LTD is \emph{manifestly causal} or, in other words, that such a representation naturally emerges from the principles of the LTD framework \cite{snowmass2020}. Regarding the set $\Sigma$, it can be obtained starting from a purely geometrical formulation \cite{Sborlini:2021owe}, similar to the well-known Cutkosky's rules \cite{Cutkosky:1960sp}.

To conclude this section, let us briefly recall a terminology introduced in Refs. \cite{Verdugo:2020kzh,Aguilera-Verdugo:2020kzc,Ramirez-Uribe:2020hes,Aguilera-Verdugo:2021nrn}. Given an $L$-loop scattering amplitude, it belongs to the Maximal Loop Topology (MLT) family if the number of propagators fulfils $n=L+1$. For the Next-to-Maximal Loop Topology (NMLT), we have $n=L+2$ and so on. In general, the N$^k$MLT family is composed by $L$-loop diagrams with $n=L+k+1$ propagators or, equivalently, it is composed by diagrams of topological order $k$.  

%20231222: OK!!
%20240104: Improved!!

%%%%%%%%%%%%%%%%%%%%%%%%%%%%%%%%%%%%%%%%%%%%%%%%%%%%%%%%%%%%%%%%%%%%%%%%%%%%%%%%%%%%%%
\section{LTD-inspired strategies for local renormalization}
\label{sec:LocalReno}
In this section, we describe different strategies to compute local renormalization counter-terms. The difference between them lies in the starting point. Whilst the first strategy makes use of the expansion around UV-propagators in Minkowski space, the second starts from the LTD causal representation directly.

%%%%%%%%%%%%%%%%%%%%%%%%%%%%%%%%%%%%%%%%%%%%%%%%%%%%%%%%%%%%%%%%%%%%%%%%%%%%%%%%%%%%%%
\subsection{UV expansion in Minkowski space}
\label{ssec:UVFelix}
As already mentioned in the introduction, given any multi-loop multi-leg amplitude, the local UV counter-terms can be built by Taylor-expanding its integrand (depending on the loop four-momenta) in the high-energy limit around the UV propagator, i.e.
\beq
G_F\left(q_{i, \uv}\right)=\frac{1}{q_{i, \uv}^2-\mu_{\uv}^2+i 0},
\label{eq:UVpropagator}
\eeq
where, for the sake of simplicity, we define $q_{i,\uv}=\ell_i$ and $\mu_{\uv}$ plays the role of a renormalization scale. The order of the expansion depends on the degree of divergence of the original amplitude in the different UV limits. The procedure is described with plenty of details in
Ref. \cite{Driencourt-Mangin:2019aix}, extending previous techniques successfully applied at one loop level \cite{Becker:2010ng,Sborlini:2016gbr,Sborlini:2016hat,Driencourt-Mangin:2019yhu}. 

To illustrate the method, let us first consider the case of the bubble integral in four space-time dimensions, i.e.
\beq
{\cal A}^{(1)} = \int \frac{d^4 k}{(k^2-m^2)\,((k-p)^2-m^2)} \, ,
\eeq
depending on the external four-momentum $p$. In this case, the superficial degree of divergence is logarithmic (i.e. equal number of powers of the momenta in the numerator and denominator), so the expansion must consider terms of, at least, order $1/|k|^4$. This leads to
\beqn
\nn \frac{1}{(k-p)^2-m^2} &=& \frac{1}{k^2-\mu_{\uv}^2}\left\{1+\frac{2 k \cdot p}{k^2-\mu_{\uv}^2} \right.
\\ \nn &-& \left. \frac{p^2-m^2+\mu_{\uv}^2}{k^2-\mu_{\uv}^2}+\frac{(2 k \cdot p)^2}{\left(k^2-\mu_{\uv}^2\right)^2}\right\} 
\\ &+& {\cal O}\left(\frac{1}{|k|^7}\right) \, ,
\eeqn
where sub-leading terms are also included. Then, the counter-term is written as
\beqn
\nn \mathcal{A}^{(1)}_{\uv} &=& \frac{1}{(k^2-\mu_{\uv}^2)^2}\left\{1+\frac{2 k \cdot p - p^2}{k^2-\mu_{\uv}^2} \right.
\\ &+& \left. 2\, \frac{m^2-\mu_{\uv}^2}{k^2-\mu_{\uv}^2}\right\} \, .
\eeqn
Therefore, the renormalized amplitude $\mathcal{A}^{(1)}_R=\mathcal{A}^{(1)}-\mathcal{A}^{(1)}_\uv$ converges, as it goes as $1/|k|^6$ in the UV limit.

The one loop case is rather simple, since the UV diverging region is well identified, even when working in the Minkowski space (i.e. directly from the Feynman representation). So, as a more refined example, let us consider a 2-loop amplitude $\mathcal{A}^{(2)}$, assumed to be free of any IR divergence, and build a set of counter-terms that render it integrable in the UV limit. The divergences of the amplitude appear in two different configurations:
\begin{enumerate}
\item when one of the loop four-momenta $\ell_1$ or $\ell_2$ tend to infinity while the other is fixed (i.e. the so-called \emph{simple UV limit}), 
\item or when both simultaneously tend to infinity (\emph{double UV limit}). 
\end{enumerate}
Thus, we need to find local counter-terms in each of these UV regions. Consequently, the total counter-term $\mathcal{A}_{\uv}^{(2)}$ is given by
\beq             
\mathcal{A}_{\uv}^{(2)}=\mathcal{A}_{\uv,1}^{(2)}+\mathcal{A}_{\uv,2}^{(2)}+\mathcal{A}_{\uv^2}^{(2)} \, ,
\label{eq:CountertermCompleto}
\eeq
where the first two terms on the r.h.s. correspond to the simple UV limit, and the last term is obtained from the double UV limit. 

In the simple UV limit, $\ell_j \rightarrow \infty$, we need first to consider the following replacements:
\beqn
\nn S_{\uv,j} : \left\{\ell_j^2 \mid \ell_j \cdot k_i\right\} &\rightarrow \left\{\lambda^2 q_{j, \uv}^2+\left(1-\lambda^2\right) \mu_{\uv}^2 \right. 
\\ & \left. \mid \lambda q_{j, \uv} \cdot k_i\right\} \, .
\label{eq:ReemplazoSIMPLE}
\eeqn
After the replacements, the expansion in $\lambda \rightarrow \infty$ to logarithmic order (denoted by the operator $L_\lambda$) allows to extract the divergent part of the integral. Therefore, the calculation of the counter-term corresponding to the loop four-momentum $\ell_j$ can be written: 
\beq
\mathcal{A}_{\uv,j}=L_\lambda\left(\left.\mathcal{A}\right|_{\mathcal{S}_{\uv,j}}\right) \, .
\label{eq:CountertermSIMPLE}
\eeq
It is worth noticing that the highest order in $\lambda$ depends on the particular expression that we are trying to renormalize: it has to be high enough to cancel all the non-integrable terms in the UV limit. Still, removing the divergent parts in the individual UV limits is not enough to guarantee integrability: the double UV limit $|\ell_i|, |\ell_j| \rightarrow \infty$ has also to be considered to cancel overlapping singularitites. The corresponding replacement is:
\beqn
\nn & \mathcal{S}_{\uv^2} : \left\{\ell_j^2 |\ell_j \cdot \ell_k | \ell_j \cdot k_i\right\} \rightarrow 
\\ \nn & \quad  \left\{\lambda^2 q_{j, \uv}^2+\left(1-\lambda^2\right) \mu_{\uv}^2 |\lambda^2 q_{j, \uv} \cdot q_{k, \uv} \right.
\\ & \left. + \left(1-\lambda^2\right) \mu_{\uv}^2 / 2 | \lambda q_{j, \uv} \cdot k_i \right\}.
\label{eq:ReemplazoDOBLE}
\eeqn
This replacement and the subsequent expansion have to be applied to the original amplitude without the simple UV divergences, namely $\mathcal{A}-\sum_{k=1}^L \mathcal{A}_{\uv,k}$. In fact, the simple UV counter-terms $\mathcal{A}_{\uv,k}$ could be divergent in the double UV limit, so these additional divergences must also be removed by the double limit counter-term. By applying the $\lambda$ expansion to logarithmic order, we get
\beq
\mathcal{A}_{\uv^2}=L_\lambda\left(\left.\left(\mathcal{A}-\sum_{j=1,12} \mathcal{A}_{ \uv,j}\right)\right|_{\mathcal{S}_{\uv^2}}\right) .
\label{eq:CountertermDOBLE}
\eeq
This strategy can be, in principle, generalized to higher-loop orders. For instance, when the diagram contains three loops, it is necessary to calculate triple limit counter-terms. The associated replacement is applied to every different possible combination of loop four-momenta that simultaneously tend to infinity, giving $\binom{L}{3}$ triple counter-terms. The replacements and expansion in $\lambda$ are applied to the original amplitude from which simple and double limit counter-terms are already subtracted. That way, the divergences are suppressed in all the possible limits until the integral converges. Still, new overlapping UV-divergent structures might appear, and this procedure could require to subtract more terms. These potential limitations are the main motivation to explore alternative frameworks.

%20231228: Finished
%20240101: Re-checked!
%20240217: OK!

%%%%%%%%%%%%%%%%%%%%%%%%%%%%%%%%%%%%%%%%%%%%%%%%%%%%%%%%%%%%%%%%%%%%%%%%%%%%%%%%%%%%%%
\subsection{Causal LTD approach}
\label{ssec:CausalLTD}
Other way to build local UV counter-terms consists in applying the UV expansion directly to the causal dual representation. In this way, the counter-term depends purely on the Euclidean momenta, instead of the four-momenta defined in a Minkowski space. This has several advantages, in particular, working in Euclidean spaces allows a natural definition of distances, so that we can easily associate the high-energy limit with the large loop three-momentum region. Furthermore, the causal dual representation lacks of non-physical singularities in the integrand, which makes the numerical integration of the renormalized amplitude much more stable.

In order to develop this procedure, we start by modifying the algorithm presented in Sec. \ref{ssec:UVFelix} focusing in promoting scalar products in Minkowski space-time to on-shell energies and scalar products in Euclidean space. 

At first order in the expansion $|\vec{\ell}| \rightarrow \infty$, the new algorithm transforms the Euclidean space propagator, $1/(\vec{\ell}^2+m^2)$, into $1/(\vec{\ell}^2+\mu_{\uv}^2)$. Thus, we consider the new replacement rule for the simple UV limit:
\beqn
\nn S_{\uv,j} &:& \left\{\vec{\ell}_j^2 \mid \vec{\ell}_j \cdot \vec{k}_i\right\} 
\\ &\rightarrow&\left\{\lambda^2 \vec{\ell}_j^2-\left(1-\lambda^2\right) \mu_{\uv}^2 \mid \lambda \vec{\ell}_j \cdot \vec{k}_i\right\},
\label{eq:ReemplazoSIMPLEcausal}
\eeqn
where $|\vec{\ell}_j| \rightarrow \infty$, whilst $\vec{k}_i$ is kept fixed at a finite value. Notice that the sign in front of the term proportional to $\mu_{\uv}^2$ is changed because we are working in a Euclidean space (i.e. we removed the energy component). Regarding the double limit $|\vec{\ell}_j|,  |\vec{\ell}_k| \rightarrow \infty$, the associated replacement is
\beqn
\nn & \mathcal{S}_{\uv^2} : \left\{\vec{\ell}_j^2 |\vec{\ell}_j \cdot \vec{\ell}_k | \vec{\ell}_j \cdot \vec{k}_i\right\} \rightarrow 
\\ \nn & \left\{\lambda^2 \vec{\ell}_j^2-\left(1-\lambda^2\right) \mu_{\uv}^2 | \right. 
\\ & \left. \lambda^2 \vec{\ell}_j \cdot \vec{\ell}_k+\left(1-\lambda^2\right) \mu_{\uv}^2 / 2 |\lambda \vec{\ell}_j \cdot \vec{k}_i\right\}.
\label{eq:ReemplazoDOBLEcausal}
\eeqn
After implementing Eqs. (\ref{eq:ReemplazoSIMPLEcausal}) and (\ref{eq:ReemplazoDOBLEcausal}), the counter-term is defined by expanding the resulting expression in $\lambda$ around infinity. Analogous replacements could be defined in the simultaneous multiple UV limit, as we will explain later in this section.

At this point, we should notice that there is a crucial detail in this algorithm: the operator $L_\lambda$ should not be applied to the prefactor $1/x_{L+k}$, because it is related to the integration measure. Let us recall the causal dual representation from Eq. (\ref{eq:DualCausalRepresentation}), symbolically written as
\beq
{\cal A}_N^{(L)} = \int_{\vec{\ell}_1 \ldots \vec{\ell}_L} \, \frac{{\cal A}_{RED}^{(L)}(\{q_{i,0}^{(+)}\}_{i=1,\ldots,L+k},\{p_j\}_{j=1,\ldots,N})}{x_{L+k}} \, ,
\label{eq:DualCausalReduced1}
\eeq
and define the \emph{reduced amplitude} ${\cal A}^{(L)}_{RED}$. It is worth to mention that, while developing this framework, we noticed that Taylor-expanding $x_{L+k}$ in the limit $\lambda \to \infty$ leads to spurious divergences that ruined the convergence in the high-energy region (in other words, that prevent the fully local cancellation of UV divergences). Also, we notice that the replacements Eqs. (\ref{eq:ReemplazoSIMPLEcausal}) and (\ref{eq:ReemplazoDOBLEcausal}) can be directly applied at the level of on-shell energies, i.e. $q_{i,0}^{(+)}$, since they transform the three-momenta.

Let us illustrate the proposed technique, considering a generic scalar $L$ loop scattering amplitude (i.e. the numerator is ${\cal N}\equiv 1$). First, we take the UV limit of $n$ loop three-momenta going simultaneously to infinity; we denote by $\gamma$ the set of indices corresponding to these loop momenta. Then, let $m$ be the number of internal lines (or propagators) that depend on, at least, one of these $n$ momenta. In this way, scaling the loop three-momenta by $\lambda$ and doing a naive power counting, we have $\lambda^{3 \, n}$ in the numerator coming from $\prod_{i \in \gamma} d^3 \vec{\ell}_i$ and $\lambda^m$ in the denominator originated from the integration measure prefactor $1/x_{L+k}$. Thus, in order to define the counter-term in this simultaneous multiple UV limit, ${\cal A}_{RED}^{(L)}$ needs to be expanded in $\lambda$ keeping terms ${\cal O}(\lambda^{3\,n - m})$, besides sub-leading powers to adjust the finite pieces. For instance, in the case of MLT diagrams \cite{Verdugo:2020kzh,Aguilera-Verdugo:2020nrp}, it would be necessary to expand up to ${\cal O}(2n-1)$. For more complicated topologies, such as N$^k$MLT, the superficial degree of UV divergence is lowered both by the prefactor ($x_{L+k} \to \lambda^{m}$) and the presence of causal entangled thresholds involving several loop momenta going to infinity.

Still, with a larger number of loops, there is a larger number of possible combinations for taking these $n$ simultaneous limits: explicitly, there are $\binom{L}{n}$ multiple UV limits. Therefore, the number of loops significantly increases the complexity of the calculations. Furthermore, overlapping singularities might appear, as in the Minkowski-space expansion, which require keeping higher-order terms in $\lambda$. The origin of such singularities is the direct implementation of replacement rules Eqs. (\ref{eq:ReemplazoSIMPLEcausal})-(\ref{eq:ReemplazoDOBLEcausal}), that could introduce additional spurious dependencies of the loop momenta in the numerators: this problem is particularly evident at three-loops and beyond.

%%%%%%%%%%%%%%%%%%%%%%%%%%%%%%%%%%%%%%%%%%%%%%%%%%%%%%%%%%%%%%%%%%%%%%%%%%%%%%%%%%%%%%
\subsection{On-shell energy expansions within causal LTD}
\label{ssec:onshellsimple}
The expansion at the level of the loop three-momenta could lead not only to cumbersome expressions, but also to alter the nice structure of causal LTD representations. Still, if it does not introduce any spurious or non-causal divergence, new functional dependences besides the on-shell energies $q_{i,0}^{(+)}$ will appear. 

Let us start with the single UV limit, and consider $q_{j,0}^{(+)}$ assuming that it depends on $\vec{\ell}_i$. Using the definition of the on-shell energies and re-scaling $ \vec{\ell}_i$, we have
\beqn
&& q_{j,0}^{(+)}=\sqrt{(\vec{\ell}_i+\vec{k})^2+m_j^2} \rightarrow
    \label{eq:qi0PLUSreemplazoBIS}
\\ \nn && \sqrt{\lambda^2\left(\vec{\ell}_i^2+\mu_{\uv}^2\right)+2\lambda \, \vec{\ell_i}\cdot\vec{k}+\vec{k}^2-\mu_{\uv}^2+m_j^2} \, ,
\eeqn
where $\vec{k}\equiv\vec{q}_j - \vec{\ell}_i$. Then, if we expand in $\lambda$, we obtain
\beqn
\nn \mathcal{S'}_{\uv,i} : q_{j, 0}^{(+)} &\rightarrow& \lambda q_{i, 0, \uv}^{(+)} + \frac{\vec{\ell}_i \cdot \vec{k}}{q_{i, 0, \uv}^{(+)}} - \frac{(\vec{\ell_i}\cdot\vec{k})^2}{2 \lambda \, (q_{i, 0, \uv}^{(+)})^3}
\\ &+& \frac{\vec{k}^2+m_j^2-\mu_\uv^2}{2 \lambda \, q_{i, 0, \uv}^{(+)}} + {\cal O}(\lambda^{-2}) \, .
\label{eqRep2BIS}
\eeqn
If we keep different masses, the replacement in the case $i=j$ with $\vec{k}=0$ takes the form
\beqn
\nn \mathcal{S'}_{\uv,i} : q_{i, 0}^{(+)} &\rightarrow& \lambda q_{i, 0, \uv}^{(+)}  + \frac{m_i^2-\mu_\uv^2}{2 \lambda \, q_{i, 0, \uv}^{(+)}} + {\cal O}(\lambda^{-2}) \, ,
\label{eqRep2BISB}
\eeqn
where the sub-leading terms depending on $m_i^2-\mu_\uv^2$ ensure the local cancellation of UV singularities.

Following these ideas, it is possible to obtain a generalization for describing the multiple UV limit. Let us consider  
\beq
\vec{q}_j=\sum_{k \in \delta}\vec{\ell}_k+\vec{p}_j \, ,
\eeq
to be the three-momentum of the $j$-th internal line with mass $m_j$, where $\vec{p}_j$ is any combination of external momenta and $\delta_j$ is the set of indices of the loop momenta that $q_j$ depends on. Let $\gamma$ be the set of loop three-momenta that are going to infinity, and define
\beqn
\vec{\ell}_{j,\gamma} &=& \sum_{k \in \gamma \cap \delta_j} \vec{\ell}_k \, ,
\\ \vec{v}_j &=& \vec{q}_j - \vec{\ell}_{j,\gamma} \, .
\eeqn
With this notation, the replacement from Eq. (\ref{eq:ReemplazoDOBLEcausal}) takes the form
\beq
q_{j,0}^{(+)} \rightarrow \sqrt{\lambda^2\, (q_{\delta_j \cap \gamma, 0, \uv}^{(+)})^2+2\lambda \, \vec{\ell}_{j,\gamma}\cdot\vec{v}_j+\vec{v}_j^2-\mu_{\uv}^2+m_j^2} \, ,
\label{eq:qi0PLUSreemplazoBIS2}
\eeq
with $q_{\delta_j \cap \gamma, 0, \uv}^{(+)}=\sqrt{\vec{\ell}_{j,\gamma}^2+\mu_{\uv}^2}$. Notice that this expression is valid regardless of the number of simultaneous divergent loop three-momenta, $\rho=\#(\gamma)$. Then, if we perform a Taylor expansion and keep terms up to ${\cal O}(\lambda^{-3})$, the replacement rule $\mathcal{S'}_{\uv^\rho,\gamma}$ simplifies to:
\beqn
\nn q_{j, 0}^{(+)} &\rightarrow& \lambda q_{\delta_j \cap \gamma, 0, \uv}^{(+)} + \frac{\vec{\ell}_{j,\gamma} \cdot \vec{v}_j}{q_{\delta_j \cap \gamma, 0, \uv}^{(+)}} - \frac{(\vec{\ell}_{j,\gamma}\cdot\vec{v}_j)^2}{2 \lambda \, (q_{\delta_j \cap \gamma, 0, \uv}^{(+)})^3}
\\ \nn &+& \frac{\vec{v}_j^2+m_j^2-\mu_\uv^2}{2 \lambda \, q_{\delta_j \cap \gamma, 0, \uv}^{(+)}} +\frac{(\vec{\ell}_{j,\gamma}\cdot\vec{v}_j)^3}{2 \lambda^2 \, (q_{\delta_j \cap \gamma, 0, \uv}^{(+)})^5}
\\ \nn &-& \frac{(\vec{\ell}_{j,\gamma}\cdot\vec{v}_j)(\vec{v}_j^2+m_j^2-\mu_\uv^2)}{2 \lambda^2 \, (q_{\delta_j \cap \gamma, 0, \uv}^{(+)})^3}  
\\ \nn &-& \frac{5(\vec{\ell}_{j,\gamma}\cdot\vec{v}_j)^4}{8 \lambda^3 \, (q_{\delta_j \cap \gamma, 0, \uv}^{(+)})^7}- \frac{(\vec{v}_j^2+m_j^2-\mu_\uv^2)^2}{8 \lambda^3 \, (q_{\delta_j \cap \gamma, 0, \uv}^{(+)})^3} 
\\ &+&\frac{3(\vec{\ell}_{j,\gamma}\cdot\vec{v}_j)^2(\vec{v}_j^2+m_j^2-\mu_\uv^2)}{4 \lambda^3 \, (q_{\delta_j \cap \gamma, 0, \uv}^{(+)})^5} + {\cal O}(\lambda^{-4}) \, .
\label{eq:Rep3GENERAL}
\eeqn
At this point, we can define an iterative procedure to locally subtract all the UV-divergent contribution for any arbitrary number of loops. Given an $L$-loop reduced scattering amplitude ${\cal A}_{RED}^{(L)}$, we calculate the simple UV counter-terms,
\beq
{\cal A}^{(L)}_{RED,\uv,i_1} = L_\lambda \left({\cal A}_{RED}^{(L)}|_{{\cal S}'_{UV^1,i_1}} \right) \, ,
\eeq
with $i_1 \in \{1, \ldots, L\}$. Then, we subtract the sum of simple UV counter-terms to the reduced amplitude, defining
\beq
{\cal A}^{(L)}_{RED,1} = {\cal A}^{(L)}_{RED,0} - \sum_{i_1 = 1}^L {\cal A}^{(L)}_{RED,\uv,i_1} \, ,
\eeq
with ${\cal A}^{(L)}_{RED,0}={\cal A}^{(L)}_{RED}$ to simplify the recursive relations. The next step consists in removing the double UV singularities, what is achieved by defining the double UV counter-terms, i.e.
\beq
{\cal A}^{(L)}_{RED,\uv,\{i_1,i_2\}} = L_\lambda \left({\cal A}_{RED,1}^{(L)}|_{{\cal S}'_{UV^2,\{i_1,i_2\}}} \right) \, ,
\eeq
and summing them over all the possible couples $\{i_1,i_2\}$. We obtain 
\beq
{\cal A}^{(L)}_{RED,2} = {\cal A}^{(L)}_{RED,1} - \sum_{i_1 = 1}^L \sum_{i_2 = i_1 + 1}^L {\cal A}^{(L)}_{RED,\uv,i_1 i_2} \, .
\eeq
Repeating the procedure $L$ times, we arrive to 
\beq
{\cal A}^{(L)}_{RED,\uv} = {\cal A}^{(L)}_{RED,L-1} -  L_\lambda \left({\cal A}_{RED,L-1}^{(L)}|_{{\cal S}'_{UV^L,\{1,\ldots,L\}}} \right) \, ,
\eeq
that corresponds to the locally renormalized reduced amplitude. Notice that this procedure is rather general, and the operator $L_\lambda$ could change step by step. This is because $L_\lambda$ involves performing the $\lambda$ series expansion to different orders, to ensure that all non-integrable terms in the limit $\vec{\ell} \to \infty$ are removed. Also, we have to take into account that keeping sub-leading orders in the first loops might lead to new and more UV singular terms for the remaining loops, implying that the expansion in $\lambda$ for the subsequent iterations had to be done to higher orders.

To conclude this section, we would like to comment on the loop momenta flow dependence. Since our approach aims to a local cancellation of the UV divergences, the UV local counter-terms could eventually exhibit an \emph{explicit} dependence on the loop momenta. Still, one crucial advantage of the causal LTD representation is that it depends only on on-shell energies, i.e. $q_{i,0}^{(+)}$. So, the loop momenta dependence is hidden inside these $q_{i,0}^{(+)}$, and any explicit loop-momenta dependence will be generated from the UV-expansion of the on-shell energies. In the next section, we will show explicit local UV counter-terms, and we will see that their explicit dependence on the loop momenta flow is minimized.

%20240128: DONE!!!
%20240218: Checado hasta aqui!

%%%%%%%%%%%%%%%%%%%%%%%%%%%%%%%%%%%%%%%%%%%%%%%%%%%%%%%%%%%%%%%%%%%%%%%%%%%%%%%%%%%%%%
\section{Benchmark multi-loop examples}
\label{sec:Examples}
In this section, we present representative examples up to 3 loops and study their numerical convergence. We will rely on the simplified strategy based on on-shell energy expansions. Besides, we would like to emphasize that our motivation here is limited to show that the UV-divergences are cancelled with the proposed local renormalization counter-terms; being able to provide precise numerical predictions would require high-precision integrators and this study is out of the scope of the present research.

%%%%%%%%%%%%%%%%%%%%%%%%%%%%%%%%%%%%%%%%%%%%%%%%%%%%%%%%%%%%%%%%%%%%%%%%%%%%%%%%%%%%%%
\subsection{\textit{Sunrise} diagram with equal masses and fixed renormalization scale}
\label{ssec:2loopMLT}
\begin{figure}[h]
\centering
\includegraphics[scale=0.3]{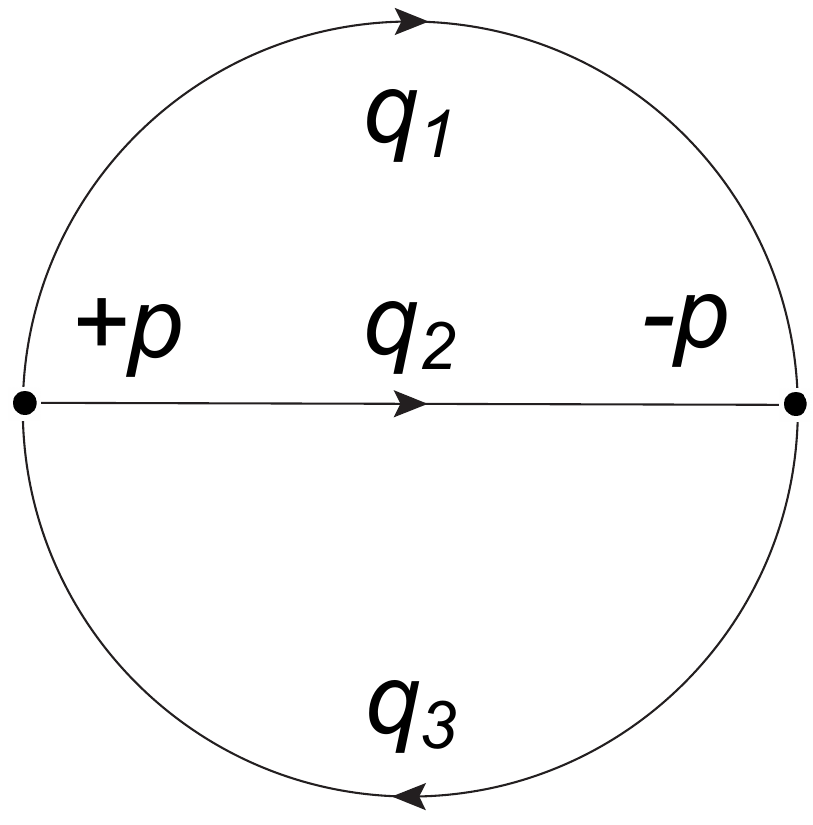}
\caption{Momenta assignation for the 2-loop MLT diagram studied, according to Eq. (\ref{eq:MomentaMLT}).}
\label{fig:MLTFig1}
\end{figure}

A simple diagram that we can use to test our local renormalization strategy is the 2-loop MLT \textit{sunrise} with an external four-momentum $p^\mu=(p_0,\vec{p})$. It is worth mentioning that, in the scalar case (i.e. ${\cal N}\equiv 1$), MLT diagrams are the most UV singular (with a superficial degree of divergence $2L-1$). For the two loop case, the causal dual representation is given by:
\beq
{\cal A}^{(2)} = \int_{\vec{\ell}_1, \vec{\ell}_2} \frac{1}{x_{3}}\left(\frac{1}{\lambda_1^+}+\frac{1}{\lambda_1^-}\right) \, ,
\label{eq:2loopMLT1}
\eeq
where 
\beq
\lambda_1^\pm = q_{1,0}^{(+)}+q_{2,0}^{(+)}+q_{3,0}^{(+)} \pm p_0 \, ,
\label{eq:ThresholdMLT1}
\eeq
are the causal thresholds associated to MLT diagrams and
\beq
q_1=\ell_1, \textbf{ } q_2=\ell_2, \textbf{ }q_3=\ell_1+\ell_2-p \, ,
\label{eq:MomentaMLT}
\eeq
is the momenta assignation, as shown in Fig. \ref{fig:MLTFig1}. The corresponding on-shell energies are given by
\beqn
q_{1,0}^{(+)} &=& \sqrt{\vec{\ell}_1^2+m_1^2} \, , 
\\ q_{2,0}^{(+)} &=& \sqrt{\vec{\ell}_2^2+m_2^2}\, , 
\\ q_{3,0}^{(+)} &=& \sqrt{\left(\vec{\ell}_1+\vec{\ell}_2-\vec{p}\right)^2+m_3^2} \, ,
\eeqn
and we set, in this subsection, $m_1=m_2=m_3=M$ as well as $\vec{p}=\vec{0}$. Therefore, the reduced amplitude is
\beqn
\nn {\cal A}^{(2)}_{RED} &=& \frac{1}{q_{1,0}^{(+)}+q_{2,0}^{(+)}+q_{3,0}^{(+)}+p_0}
\\ &+&\frac{1}{q_{1,0}^{(+)}+q_{2,0}^{(+)}+q_{3,0}^{(+)}-p_0} \, .
\label{eq:2loopMLT2}
\eeqn
This diagram depends on two loop momenta $\vec{\ell}_1$ and $\vec{\ell}_2$, so the potential UV-divergent regions to be considered are: 
$|\vec{\ell}_1| \rightarrow \infty$ and $|\vec{\ell}_2| \rightarrow \infty$, as well as the simultaneous limit $|\vec{\ell}_1|, |\vec{\ell}_2| \rightarrow \infty$.

%%%%%%%%%%%%%%%%%%%%%%%
%%%%%%%%%%%%%%%%%%%%%%

In the case $|\vec{\ell}_1| \rightarrow \infty$, Eq. (\ref{eqRep2BIS}) yields the replacement $q_{1,0}^{(+)} \rightarrow \lambda \, q_{1,0,\uv}^{(+)}$ keeping only the leading order in $\lambda$, while $q_{3,0}^{(+)} \rightarrow \lambda \, q_{1,0,\uv}^{(+)}$ due to the dependence of $\vec{q}_3$ on $\vec{\ell}_1$. After applying these transformations to the reduced amplitude, we get
\beqn
   \nn {\cal A}_{RED}^{(2)} \mid_{ \mathcal{S'}_{\uv,1}} &=& \frac{1}{2\, \lambda q_{1,0,\uv}^{(+)}+q_{2,0}^{(+)}+p_0}
   \\ &+& \frac{1}{2\, \lambda q_{1,0,\uv}^{(+)}+q_{2,0}^{(+)}-p_0},
\eeqn
to which the expansion $\lambda \rightarrow \infty$ is applied up to order 1, in accordance with the formula found in Sec. \ref{sec:CausalLTD}: $3n-2=1$ for $n=1$ (single UV limit) and $m=2$ (two other momenta depending on $\vec{\ell}_1$). Finally the limit $\lambda \rightarrow 1$ is taken, and the reduced part of the counter-term is given by
\beq
    {\cal A}^{(2)}_{RED,\uv,1}= \frac{1}{q_{1,0,\uv}^{(+)}} \, ,
    \label{eq:A2REDUV_MLT2}
\eeq
meaning that the total counter-term in the limit $|\vec{\ell}_1| \rightarrow \infty$ is 
\beq
    {\cal A}_{\uv,1}^{(2)}=\int_{\vec{\ell}_1, \vec{\ell}_2} \, \frac{1}{x_3 \, q_{1,0,\uv}^{(+)}} \, .
\eeq
In the case $|\vec{\ell}_2| \rightarrow \infty$, we can exploit the exchange symmetry of the indices $1 \longleftrightarrow 2$. This leads directly to the counter-term 
\beq
    {\cal A}_{\uv,2}^{(2)}=\int_{\vec{\ell}_1, \vec{\ell}_2} \frac{1}{x_3 \, q_{2,0,\uv}^{(+)}} \, .
\eeq

For the double UV limit, we first define
\beq
(\mathcal{A}_{RED}^{(2)})^\prime = \mathcal{A}_{RED}^{(2)}-\mathcal{A}_{RED,\uv, 1}^{(2)}-\mathcal{A}_{RED,\uv, 2}^{(2)} \, ,
\eeq
that corresponds to the original reduced amplitude after subtracting the single UV counter-terms. Using Eq. (\ref{eq:Rep3GENERAL}), we have
\beqn
q_{1,0}^{(+)} &\rightarrow& \lambda \, q_{1,0, \uv}^{(+)} \, ,
\\ q_{2,0}^{(+)} &\rightarrow& \lambda \, q_{2,0, \uv}^{(+)} \, ,
\\ q_{3,0}^{(+)} &\rightarrow& \lambda \, q_{12,0, \uv}^{(+)} \, ,
\eeqn
with 
\beq
q_{12,0,\uv}^{(+)}=\sqrt{\left(\vec{\ell}_1+\vec{\ell}_2\right)^2+\mu_{\uv}^2} \, ,
\eeq
and the condition $\mu_\uv \equiv M$. These replacements leads to
\beqn
\nn && (\mathcal{A}_{RED}^{(2)})^{\prime} \mid_{ \mathcal{S'}_{\uv^2}} = \frac{1}{\lambda\left(q_{1,0, \uv}^{(+)}+q_{2,0, \uv}^{(+)}+q_{12,0, \uv}^{(+)}\right)+p_0} 
\\ && + \frac{1}{\lambda\left(q_{1,0, \uv}^{(+)}+q_{2,0, \uv}^{(+)}+q_{12,0, \uv}^{(+)}\right)-p_0} \, .
\eeqn
In this case, the expansion in $\lambda$ is carried out to order three. After taking the limit $\lambda \rightarrow 1$ and restoring the prefactor, the counter-term of the double UV limit is 
\beqn
\nn \mathcal{A}_{\uv,12}^{(2)} &=& \int_{\vec{\ell}_1,\vec{\ell}_2} \frac{1}{x_3}
\\ \nn &\times& \left(\frac{2 p_0^2}{\left(q_{1,0, \uv}^{(+)}+q_{2,0, \uv}^{(+)}+q_{12,0, \uv}^{(+)}\right)^3} \right. 
\\ \nn&+& \left. \frac{2}{q_{1,0, \uv}^{(+)}+q_{2,0, \uv}^{(+)}+q_{12,0, \uv}^{(+)}} \right.
\\ &-& \left. \frac{1}{q_{1,0, \uv}^{(+)}}-\frac{1}{q_{2,0, \uv}^{(+)}}\right) .
\eeqn
Finally, the total counter-term for the $\textit{sunrise}$ diagram is obtained summing the counter-terms for the three limits:
\beqn
\nn \mathcal{A}^{(2)}_{\uv} &=& \mathcal{A}^{(2)}_{\uv,1}+\mathcal{A}^{(2)}_{\uv,2}+\mathcal{A}^{(2)}_{\uv,12} 
\\ &=& \int_{\vec{\ell}_1,\vec{\ell}_2} \frac{2}{x_3} \left( \frac{p_0^2}{Q_\uv^3} + \frac{1}{Q_\uv} \right) \, ,
\label{eq:FullCountertermSUNRISE}
\eeqn
where we defined $Q_\uv=q_{1,0, \uv}^{(+)}+q_{2,0,\uv}^{(+)}+q_{12,0, \uv}^{(+)}$, namely the UV-version of the causal threshold present in this MLT diagram. By Taylor expanding, it can be proved that the renormalized amplitude ${\cal A}^{(2)}_R={\cal A}^{(2)}-{\cal A}^{(2)}_{\uv}$ converges since the divergent orders are exactly cancelled in both the simple and the double UV limit.

This example shows the importance of applying the replacement and subsequent Taylor expansion to $({\cal A}^{(2)}_{RED})^\prime$ instead of ${\cal A}^{(2)}_{RED}$. This is because the simple UV counter-terms ${\cal A}^{(2)}_{RED,\uv,1}$ and ${\cal A}^{(2)}_{RED,\uv,2}$ add spurious divergences in the double UV limit, which are then subtracted by ${\cal A}^{(2)}_{RED,\uv,12}$. In this particular example, the counter-terms for the simple UV limits are completely eliminated by ${\cal A}^{(2)}_{RED,\uv,12}$ and they do not appear in the complete UV counter-term. Besides this, it is important to work with the reduced amplitudes, since the expansion of the prefactor associated to the phase-space measure would introduce additional contributions that prevent a local UV renormalization, unless extra terms are introduced. It is also worth noticing that, in the particular case of $\mathcal{A}^{(2)}_{\uv}$, all the dependencies on the masses and renormalization scale are embodied within the on-shell energies.

%%%%%%%%%%%%%%%%%%%%%%%%%%%%%%
\begin{figure}[h]
\centering
\includegraphics[scale=0.95]{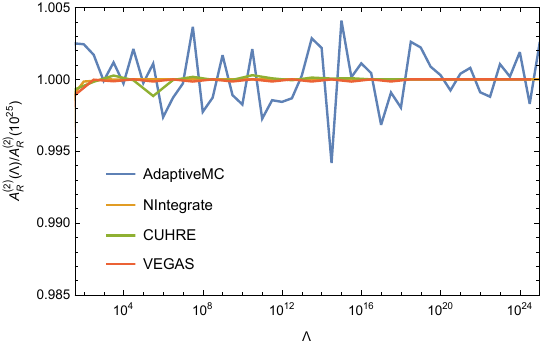}
\caption{Analysis of the numerical convergence for the \textit{sunrise} with equal masses and $\mu_\uv = M$, as a function of the cut-off $\Lambda$. The results are normalized to ${\cal A}^{(2)}_{R}(\Lambda=10^{25})$, using $M=4/10$, $p_0=2/10$ and neglecting units. Four different setups are considered for performing the numerical integration, as described in the text.}
\label{fig:MLT}
\end{figure}
%%%%%%%%%%%%%%%%%%%%%%%%%%%%%%

Another interesting remark is that the replacements $\mathcal{S}'_{\uv}$, as given in Eqs. (\ref{eqRep2BIS}) and (\ref{eq:Rep3GENERAL}) keeping only the leading order in $\lambda$, are defined by the Taylor expansion of $\mathcal{S}_{\uv}$ at first order in $\lambda$. But, for obtaining the double UV counter-term $\mathcal{A}_{RED,\uv,12}^{(2)}$, an expansion in $\lambda$ up to order three is carried out. Therefore, the sub-leading terms of the original replacement $\mathcal{S}_{\uv^2}$, absent in $\mathcal{S}'_{\uv^2}$, are missing when this new substitution is applied. The convergence of the renormalized amplitude, even without these missing terms (i.e. just taking into account the leading term of the replacement) is very smooth.

%20240110: Finished!

%%%%%%%%%%%%%%%%%%%%%%%%%%%%%%%%%%%%%%%%%%%%%%%%%%%%%%%%%%%%%%%%%%%%%%%%%%%%%%%%%%%%%%
\subsubsection{Numerical integration}
\label{sssec:NumericalIntegration}
After proving that the renormalized amplitude ${\cal A}_R^{(2)}$ is integrable in the UV region by construction, we test the numerical stability of the formalism in $d=4$ space-time dimensions. For doing so, we use first spherical coordinates to parametrize the loop three-momenta, i.e.
\beq
\vec{\ell_i} = \ell_i \,  \{\sin(\theta_i)\cos(\phi),\sin(\theta_i)\sin(\phi),\cos(\theta_i)\} \, ,
\eeq
with $\ell_i \in (0,\infty)$. Then, we compactify the integration domain by changing variables according to
\beq
\ell_i = \frac{x_i}{1-x_i} \, ,
\label{eq:Compactify}
\eeq
where $x_i \in (0,1)$ and the integration measure from Eq. (\ref{eq:IntegrationMeasure1}) reads
\beq
\int_{\vec{\ell_i}} = \int_0^1  dx_i \, \int_0^\pi d\theta_i \,  \int_0^{2\pi} d\phi_i \, \frac{\sin(\theta_i) \, x_i^2}{8\pi^3\, (1-x_i)^4} \, .
\eeq
The UV limit is reached when $x_i \to 1$. Hence, for studying the quality of the convergence of the locally renormalized amplitude, we introduce a cut-off energy, $\Lambda$, such that $\ell_i < \Lambda$ for all the loop three-momenta (in this example, $i=\{1,2\}$) and numerically evaluated ${\cal A}_R^{(2)}$ for different values of $\Lambda$. Switching to $x_i$, this means that the upper limit should set to
\beq
x_i^{\rm MAX} = \frac{\Lambda}{1+\Lambda} \, .
\eeq
If the cancellation of UV divergences is stable, we expect ${\cal A}_R^{(2)}$ to converge as $\Lambda \to \infty$ (or $x_i \to 1$).

%%%%%%%%%%%%%%%%%%%%%%%%%%%%%%
\begin{figure}[htb]
\centering
\includegraphics[scale=0.95]{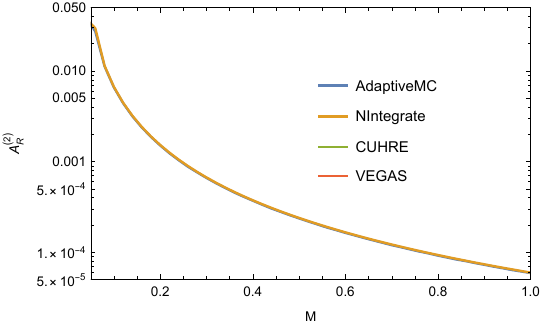}
\caption{Analysis of the mass dependence of the renormalized amplitude ${\cal A}^{(2)}_{R}$, with a fixed cut-off $\Lambda=10^{25}$ and $p_0=2/10$. Four different setups are considered for performing the numerical integration, as described in the text.}
\label{fig:MLTmasas}
\end{figure}
%%%%%%%%%%%%%%%%%%%%%%%%%%%%%%

In Fig. \ref{fig:MLT}, we present the results for a fully numerical integration of ${\cal A}_{R}^{(2)}$ as a function of the cut-off $\Lambda$. We work in arbitrary units, fixing $p_0=0.2$ for the external momenta, $M=0.4$ for the three internal lines and $\mu_{\uv}=0.4$ as the renormalization scale. For testing purposes within \texttt{Mathematica}, we used four different setups: \texttt{NIntegrate} with \texttt{AdaptiveMonteCarlo}, the default \texttt{NIntegrate}, and \texttt{Cuhre} and \texttt{Vegas} from \texttt{Cuba} library \cite{Hahn:2004fe}. The renormalized amplitude increases as higher-energy contributions are taken into account, stabilizing very fast and converging to ${\cal A}_R^{(2)}=(372 \pm 2) \, \times 10^{-6}$. This value is obtained from the average of the integral within the four different scenarios at $
\Lambda = 10^{25}$, and leads to a relative error estimation of ${\cal O}(1 \, \%)$. Also, we consider the error estimation provided by \texttt{Vegas}, which is fully compatible with the statistical fluctuations among the different integration methods: in concrete, it returns ${\cal O}(0.4 \, \%)$, on average over the whole range of $\Lambda$.

Finally, we present in Fig. \ref{fig:MLTmasas} a plot showing the value of the numerical calculation of the renormalized amplitude for different values of $M$. The cut-off energy was $\Lambda = 10^{25}$ and the energy of the external particle was fixed to $p_0=2/10$. We considered different computational setups to test the numerical stability and the smoothness of the mass dependence. In fact, we can see that they are in perfect agreement with each other. For $M>0.06$, the relative error estimation provided by \texttt{Vegas} is ${\cal O}(0.4 \, \%)$ and the relative differences comparing the four different methods is ${\cal O}(1.5 \, \%)$. The largest deviations arise in the limit $M\to0$, where IR singularities appear. In that region, \texttt{Vegas} leads to an error of ${\cal O}(15-80 \, \%)$ with 500000 integrand evaluations.

%20240117: DONE
%20240430: REVISADO y ACTUALIZADO!

%%%%%%%%%%%%%%%%%%%%%%%%%%%%%%%%%%%%%%%%%%%%%%%%%%%%%%%%%%%%%%%%%%%%%%%%%%%%%%%%%%%%%%
\subsection{Generic \textit{Sunrise} diagram}
\label{ssec:2loopMLTfree}
For a generic sunrise diagram, truncating the expansion in $\lambda$ and retaining only the leading order within the replacement rules given is not enough to locally cancel the UV divergences. For instance, non-trivial terms involving differences of the masses appear and prevent integrability in the double UV limit. Thus, we need to keep sub-leading terms in $\lambda$ within the on-shell energies. If we do so, the total counter-term obtained for the generic sunrise diagram (i.e. a 2-loop MLT) is given by
\beqn
\nn \mathcal{A}^{(2)}_{\uv} &=& \int_{\vec{\ell}_1,\vec{\ell}_2} \frac{1}{x_3} \,  \left\{ \frac{2}{Q_\uv}+\frac{1}{Q_\uv^2}\left[ 2\, \vec{\ell}_{12} \cdot \vec{p} + \frac{m_1^2}{q_{1,0,\uv}^{(+)}} \vphantom{\frac{m_1^2}{q_{1,0,\uv}}} \right. \right.
\\ \nn &+& \left. \left. \frac{m_2^2}{q_{2,0,\uv}^{(+)}}-\mu_{\uv}^2\left(\frac{1}{q_{1,0,\uv}^{(+)}}+\frac{1}{q_{2,0,\uv}^{(+)}}\right)\right]   \right.
\\ \nn &+& \left. \frac{2p_0^2}{Q_\uv^3} + \frac{1}{Q_\uv^2 (q_{12,0, \uv}^{(+)})^3}\left[\vphantom{\frac{m_1^2}{q_{1,0,\uv}^{(+)}}} \vec{\ell}_{12}^2\, (m_3^2+\vec{p}^2) \right. \right.
\\ \nn &-& \left. \left. \left(1+\frac{2\,q_{12,0, \uv}^{(+)}}{Q_\uv}\right)(\vec{\ell}_{12} \cdot \vec{p})^2 - \mu_\uv^4\right.  \right.
\\ &+& \left. \left. \mu_\uv^2(m_3^2+\vec{p}^2-\vec{\ell}_{12}^2)  
 \vphantom{\frac{m_1^2}{q_{1,0,\uv}^{(+)}}} \right] \vphantom{\frac{m_1^2}{q_{1,0,\uv}^{(+)}}} \right\} \, .
\label{eq:CountertermUnequalMassSUNRISE}
\eeqn
Notice that the structure in the numerator is far more complex than the one found in the equal-mass case. Still, if we set $m_i\equiv M$, $\vec{p}=0$ and $\mu_\uv=M$, Eq. (\ref{eq:CountertermUnequalMassSUNRISE}) reduces to Eq. (\ref{eq:FullCountertermSUNRISE}). Also, we notice that the explicit loop momenta flow dependence is rather minimal and only manifests in the numerators: the denominators solely contain combinations of on-shell energies.

%%%%%%%%%%%%%%%%%%%%%%%%%%%%%%
\begin{figure}[!h]
\centering
\includegraphics[scale=0.95]{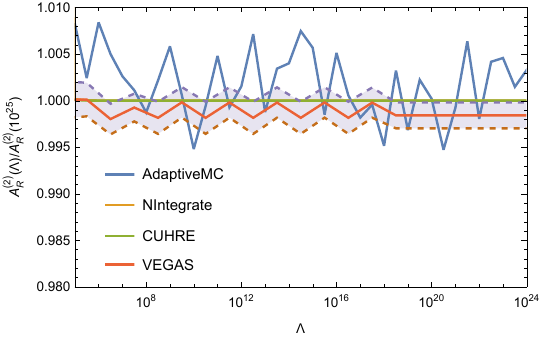}
\caption{Analysis of the numerical convergence for the \textit{sunrise} with different masses, as a function of the cut-off $\Lambda$. The results are normalized to ${\cal A}^{(2)}_{R}(\Lambda=10^{25})$, using $m_1=3/10$, $m_2=m_3=5/10$, $\mu_\uv=1$, $p_0=2/10$ and neglecting units. Four different setups are considered for performing the numerical integration, as described in the text. The purple error band is generated using the output of \texttt{Vegas}.}
\label{fig:MLT2}
\end{figure}
%%%%%%%%%%%%%%%%%%%%%%%%%%%%%%

%%%%%%%%%%%%%%%%%%%%%%%%%%%%%%
\begin{figure}[htb]
\centering
\includegraphics[scale=0.95]{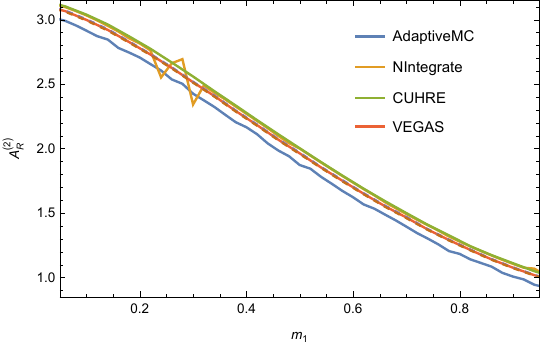}
\caption{Analysis of the mass dependence of the renormalized amplitude ${\cal A}^{(2)}_{R}$, with a fixed cut-off $\Lambda=10^{25}$. We set $m_2=m_3=5/10$, $p_0=2/10$ and $\mu_\uv=1$. Three different setups are considered for performing the numerical integration, as described in the text. The error band (purple) is provided by \texttt{Vegas}, although it can not be appreciated in the plot since it is very small, i.e. ${\cal O}(0.5 \, \%)$.}
\label{fig:MLT2masas}
\end{figure}
%%%%%%%%%%%%%%%%%%%%%%%%%%%%%%

Once the counter-term was analytically computed, we proceed to test numerically the quality of the convergence. In first place, we set $\vec{p}=0$, which is still a rather general case (essentially, this covers any time-like or null vector $p^\mu$ when $p_0$ is real). In Fig. \ref{fig:MLT2}, we consider $m_1=3/10$, $m_2=m_3=5/10$, $p_0=2/10$ and $\mu_\uv=1$, in the four scenarios described in Sec. \ref{sssec:NumericalIntegration}. The renormalized amplitude increases as higher-energy contributions are taken into account, stabilizing very fast and converging to ${\cal A}_R^{(2)}=2.46 \pm 0.09$. This value is obtained from the average of the integral within the four different scenarios at $
\Lambda = 10^{25}$. Regarding the error estimation, we obtain a relative error of ${\cal O}(7.8 \, \%)$ from the comparison among methods, whilst \texttt{Vegas} leads to ${\cal O}(0.4\, \%)$. At this point, we will consider as default estimator during the rest of the work the error provided by \texttt{Vegas}, since the largest discrepancies come from \texttt{Cuhre} and the default setup of \texttt{NIntegrate}. We found that \texttt{Cuhre} largely overestimates the error, leading to ${\cal O}(60 \, \%)$, although the difference among the central values w.r.t. \texttt{Vegas} is ${\cal O}(4 \, \%)$. 

%%%%%%%%%%%%%%%%%%%%%%%%%%%%%%
\begin{figure}[htb]
\centering
\includegraphics[scale=0.95]{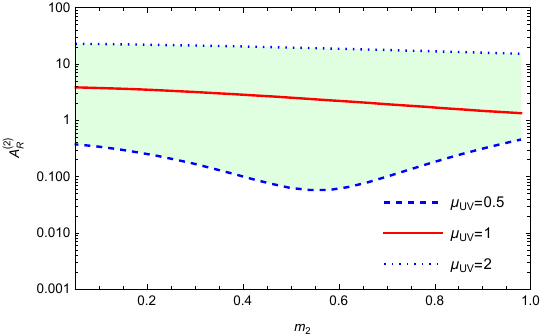}
\caption{Analysis of the mass and renormalization scale dependence $\mu_\uv$ of ${\cal A}^{(2)}_{R}$, with a fixed cut-off $\Lambda=10^{25}$ and $m_2 \in (1/50,1)$. The external energy is fixed to $p_0=2/10$, as well as $m_1=5/10$ and $m_3=3/10$. We use \texttt{Vegas} method to perform the numerical integration.}
\label{fig:MLT2renormalization}
\end{figure}
%%%%%%%%%%%%%%%%%%%%%%%%%%%%%%

Additionally, we test the stability of the numerical results when varying the values of the masses. In particular, as we show in Fig. \ref{fig:MLT2masas}, we keep $m_2=m_3=5/10$, $p_0=2/10$ and $\mu_\uv=1$ fixed, and we consider $m_1 \in (1/50,1)$. The numerical integration was performed using \texttt{NIntegrate} with \texttt{AdaptiveMonteCarlo} (blue), the default \texttt{NIntegrate} (orange), \texttt{Cuhre} (green) and \texttt{Vegas} (red). The dependence in $m_1$ is very smooth, and the error band delimited by the central values obtained with the four methods is ${\cal O}(5 \, \%)$ for the low-mass region, although reaches up to ${\cal O}(50 \, \%)$ for $m_1\approx 1$. We notice that the discrepancies between \texttt{NIntegrate} with \texttt{AdaptiveMonteCarlo} (blue) and the default \texttt{NIntegrate} (orange) are ${\cal O}(5 \, \%)$ for the whole range of $m_1$, and a similar result is obtained when comparing \texttt{Cuhre} (green) and \texttt{Vegas} (red). Again, \texttt{Cuhre} gives a largely overestimated relative error, whilst \texttt{Vegas} with 5000000 points and 10 iterations leads to ${\cal O}(0.5 \, \%)$.

Finally, we study the dependence on the renormalization scale $\mu_\uv$. For this purpose, we fix $m_1=5/10$, $m_3=3/10$ and $p_0=2/10$ (with $\vec{p}=0$, as in the previous examples shown in this section). The cut-off scale is set to $\Lambda = 10^{25}$, and we rely on \texttt{Vegas} with 5000000 points and 10 iterations to integrate the expressions. The relative error provided by this method is ${\cal O}(0.8 \, \%)$. In Fig. \ref{fig:MLT2renormalization}, we vary the value of $m_2$ within the range $(1/50,1)$, and consider three different values of $\mu_\uv$: $1/2$ (dashed blue), $1$ (red) and $2$ (dotted blue). The green region serves as a estimator of the perturbative error associated to ${\cal A}^{(2)}_R(\Lambda=10^{25})$. We observe that the counter-term defined in Eq. (\ref{eq:CountertermUnequalMassSUNRISE}) successfully cancel, at \emph{integrand level}, all the UV singular terms for arbitrary values of masses and the renormalization scale.

%20240208: Done.

%%%%%%%%%%%%%%%%%%%%%%%%%%%%%%%%%%%%%%%%%%%%%%%%%%%%%%%%%%%%%%%%%%%%%%%%%%%%%%%%%%%%%%
\subsection{3-loop MLT diagram}
\label{ssec:3loopMLT}
The next step in complexity consists in locally renormalizing a 3-loop amplitude. The simplest but most divergent scalar example is the 3-loop MLT diagram, given by
\beq
{\cal A}^{(3)} = \int_{\vec{\ell}_1, \vec{\ell}_2,\vec{\ell}_3} \frac{1}{x_{4}}\left(\frac{1}{\lambda_1^+}+\frac{1}{\lambda_1^-}\right) \, ,
\label{eq:3loopMLT1}
\eeq
with
\beq
\lambda_1^\pm = q_{1,0}^{(+)}+q_{2,0}^{(+)}+q_{3,0}^{(+)}+q_{4,0}^{(+)} \pm p_0 \, ,
\eeq
the causal threshold, which is totally analogous to the 2-loop MLT \cite{Verdugo:2020kzh}. The momenta assignation is given by
\beq
q_1=\ell_1, \textbf{ } q_2=\ell_2, \textbf{ }q_3=\ell_3, \textbf{ }q_4=\ell_1+\ell_2+\ell_3-p \, ,
\label{eq:MomentaMLT3}
\eeq
and the on-shell energies are
\beqn
q_{1,0}^{(+)} &=& \sqrt{\vec{\ell}_1^2+m_1^2} \, , 
\\ q_{2,0}^{(+)} &=& \sqrt{\vec{\ell}_2^2+m_2^2}\, , 
\\ q_{3,0}^{(+)} &=& \sqrt{\vec{\ell}_3^2+m_3^2} \, ,
\\ q_{4,0}^{(+)} &=& \sqrt{\left(\vec{\ell}_1+\vec{\ell}_2+\vec{\ell}_3-\vec{p}\right)^2+m_4^2} \, .
\eeqn
Then, the reduced amplitude is obtained from the integrand of Eq. (\ref{eq:3loopMLT1}) by removing the prefactor $1/x_4$. The procedure for computing the local counter-term is analogous to the one described for the generic sunrise, although the intermediate step expressions are more lengthy. Firstly, we need to compute the single UV counter-terms for $|\vec{\ell}_i|\to \infty$ for $i=\{1,2,3\}$, expanding the reduced amplitude in $\lambda$ and retaining up to ${\cal O}(\lambda^{-1})$ terms: these are equivalent to Eq. (\ref{eq:A2REDUV_MLT2}). Then, the double UV counter-terms require to expand up to ${\cal O}(\lambda^{-3})$, whilst the triple UV counter-terms up to ${\cal O}(\lambda^{-5})$. For the sake of completeness, the final counter-term is provided in an ancillary file \cite{ZENODO}.

%%%%%%%%%%%%%%%%%%%%%%%%%%%%%%
\begin{figure}[!h]
\centering
\includegraphics[scale=0.95]{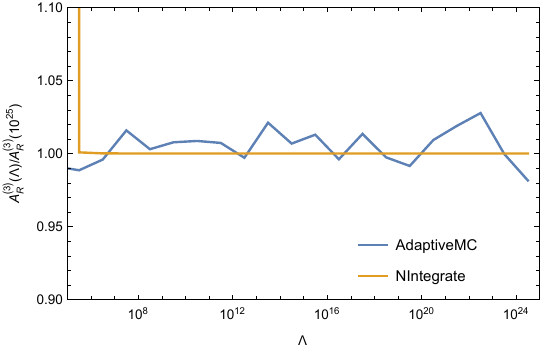}
\caption{Analysis of the numerical convergence for the 3-loop MLT diagram with different masses, as a function of the cut-off $\Lambda$. The results are normalized to ${\cal A}^{(3)}_{R}(\Lambda=10^{25})$, using $m_1=4/10$, $m_2=m_3=m_4=6/10$, $\mu_\uv=1$, $p_0=2/10$ and neglecting units.}
\label{fig:MLT3}
\end{figure}
%%%%%%%%%%%%%%%%%%%%%%%%%%%%%%

After deriving the counter-term, we proceed to test the numerical cancellation of non-integrable contributions in all the UV limits. The locally renormalized amplitude behaves as $1/|\vec{\ell}|^4$, $1/|\vec{\ell}|^7$ and $1/|\vec{\ell}|^{10}$ in the single, double and triple UV limits, respectively: this implies that it is integrable in the whole UV region.

%%%%%%%%%%%%%%%%%%%%%%%%%%%%%%
\begin{figure}[htb]
\centering
\includegraphics[scale=0.95]{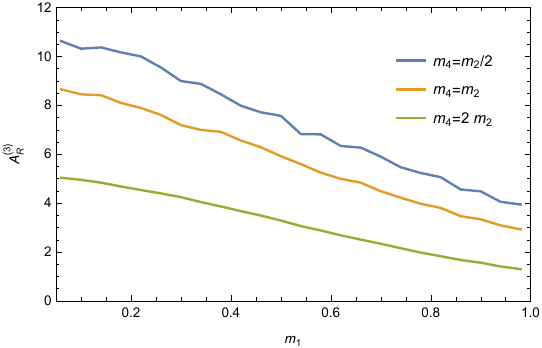}
\caption{Analysis of the mass dependence of the renormalized amplitude ${\cal A}^{(3)}_{R}$, with a fixed cut-off $\Lambda=10^{25}$. We set $m_2=m_3=5/10$, $p_0=2/10$ and $\mu_\uv=1$. Then, we vary $m_1 \in (1/48,1)$ and consider three different values of $m_4$: $m_4=m_2 / 2$ (blue), $m_4=m_2$ (orange) and $m_4=2m_2$ (green).}
\label{fig:MLT3masas}
\end{figure}
%%%%%%%%%%%%%%%%%%%%%%%%%%%%%%

Following with the tests, we study the convergence of the numerical integration for increasing values of the UV cut-off. After setting $\vec{p}=0$, we consider $m_1=4/10$, $m_2=m_3=m_4=6/10$, $p_0=2/10$ and $\mu_\uv=1$, using two different methods within \texttt{NIntegrate}: \texttt{AdaptiveMonteCarlo} (blue line) and default configuration (orange line). For this particular case, even if the local cancellation of UV singularities is guaranteed, the numerical precision required to converge in a reasonable amount of time exceed the limit of double-precision available within \texttt{Cuba} library. Thus, in this subsection, we rely only on \texttt{NIntegrate} since it allows to perform the calculations with more than 100-digits of precision. The results are shown in Fig. \ref{fig:MLT3}. The relative error associated to \texttt{NIntegrate} with \texttt{AdaptiveMonteCarlo} is ${\cal O}(2 - 5 \, \%)$, using 1000000 points, whilst the default configuration provides an overestimation of ${\cal O}(100 \, \%)$: in any case, both methods lead to compatible results within the error bands. The renormalized amplitude tends to stabilize very fast, already reaching the asymptotic value for $\Lambda=10^{5}$ with the default \texttt{NIntegrate} method. Still, MonteCarlo based methods tends to oscillate more and the convergence occurs within a band of ${\cal O}(10 \, \%)$. By averaging over these two methods, the renormalized amplitude is ${\cal A}_R^{(3)}(\Lambda=10^{25})=3.5 \pm 1.3$. Notice that the error is larger than the one found for the generic sunrise diagram; this could be further reduced by increasing the number of evaluations within the integrators.

%%%%%%%%%%%%%%%%%%%%%%%%%%%%%%
\begin{figure}[htb]
\centering
\includegraphics[scale=0.95]{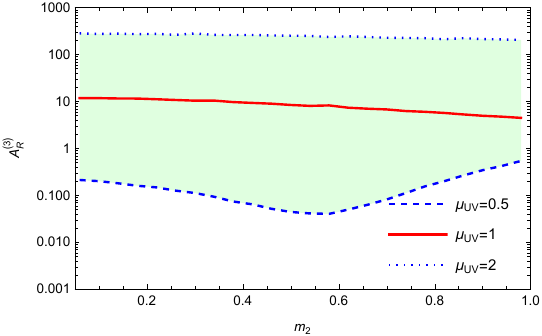}
\caption{Analysis of the mass and renormalization scale dependence $\mu_\uv$ of ${\cal A}^{(3)}_{R}$, with a fixed cut-off $\Lambda=10^{25}$ and $m_2 \in (1/50,1)$. The external energy is fixed to $p_0=2/10$, as well as $m_1=5/10$ and $m_3=m_4=3/10$. We use \texttt{NIntegrate} with \texttt{AdaptiveMonteCarlo} method to perform the numerical integration.}
\label{fig:MLT3renormalization}
\end{figure}
%%%%%%%%%%%%%%%%%%%%%%%%%%%%%%

Again, we test the stability of the numerical results when varying the values of the masses. In particular, as we show in Fig. \ref{fig:MLT3masas}, we keep $m_2=m_3=5/10$, $p_0=2/10$ and $\mu_\uv=1$ fixed. We consider $m_1 \in (1/48,1)$, and three different scenarios: $m_4=m_2/2$ (blue), $m_4=m_2$ (orange) and $m_4=2m_2$ (green). The numerical integration was performed using \texttt{NIntegrate} with \texttt{AdaptiveMonteCarlo}. We tested other numerical integrators but they were less efficient in converging. As we saw in the 2-loop MLT case, the dependence in $m_1$ is very smooth, as well as the transition for different values of $m_4$. Furthermore, the results follow the expected behaviour: the heavier the particles, the less probable their production or their presence within the loop.

To conclude, we study the dependence on the renormalization scale $\mu_\uv$. We fix $m_1=5/10$, $m_3=m_4=3/10$ and $p_0=2/10$ (with $\vec{p}=0$), together with $\Lambda = 10^{25}$. In Fig. \ref{fig:MLT3renormalization}, we vary the value of $m_2$ within the range $(1/48,1)$, and consider three different values of $\mu_\uv$: $1/2$ (dashed blue), $1$ (red) and $2$ (dotted blue). As expected, the local counter-term cancel all the UV singular terms for arbitrary values of masses and the renormalization scale. Still, we can appreciate that the dependence on $\mu_\uv$ is very strong: the results change roughly two orders of magnitude when modifying the scale by a factor two up or down. This suggest a very strong divergent behaviour in the UV, pointing to the correctness of the proposed local counter-terms (i.e. if they were wrong, the result would diverge wildly).

%20240214: Done!!!!
%20240219: RE-CHECKED!!

%%%%%%%%%%%%%%%%%%%%%%%%%%%%%%
\begin{figure}[h]
\centering
\includegraphics[scale=0.6]{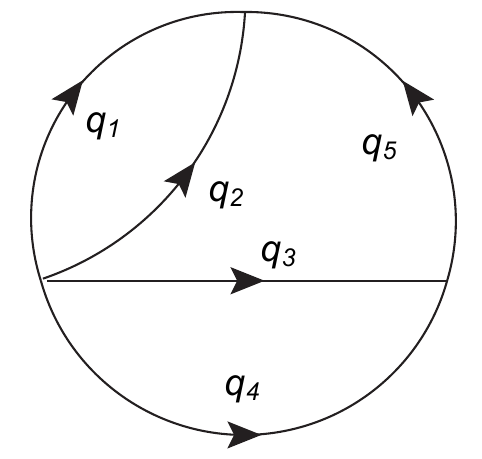}
\caption{Momenta assignation for the 3-loop NMLT diagram according to Eq. (\ref{eq:MomentaNMLT}).}
\label{fig:NMLTFig1}
\end{figure}
%%%%%%%%%%%%%%%%%%%%%%%%%%%%%%

%%%%%%%%%%%%%%%%%%%%%%%%%%%%%%%%%%%%%%%%%%%%%%%%%%%%%%%%%%%%%%%%%%%%%%%%%%%%%%%%%%%%%%
\subsection{3-loop NMLT diagram}
\label{ssec:3loopNMLT}
The final benchmark example reported in this work is a 3-loop NMLT diagram without any external momenta (i.e. a vacuum diagram). The corresponding reduced amplitude is given by
\beq
 {\cal A}^{(3)}_{RED} =\frac{1}{\lambda_1 \, \lambda_2} + \frac{1}{\lambda_2 \, \lambda_3} + \frac{1}{\lambda_3 \, \lambda_1} \, ,
\label{eq:3loopMLT2}
\eeq
where
\beqn
\lambda_1 &=& q_{1,0}^{(+)} +q_{2,0}^{(+)}+q_{3,0}^{(+)}+q_{4,0}^{(+)} \, ,
\\ \lambda_2 &=& q_{1,0}^{(+)} +q_{2,0}^{(+)}+q_{5,0}^{(+)} \, ,
\\ \lambda_3 &=& q_{3,0}^{(+)} +q_{4,0}^{(+)}+q_{5,0}^{(+)} \, ,
\eeqn
are the causal thresholds associated to the vacuum NMLT topology and
\beqn
\nn && q_1=\ell_1, \textbf{ } q_2=\ell_2,
\textbf{ }q_3=\ell_3,
\\ && q_4=-\ell_1-\ell_2-\ell_3,\textbf{ }q_5=-\ell_1-\ell_2 \, ,
\label{eq:MomentaNMLT}
\eeqn
is the momenta assignation, as shown in Fig. \ref{fig:NMLTFig1}. The explicit expression of the on-shell energies is given by
\beqn
q_{1,0}^{(+)} &=& \sqrt{\vec{\ell}_1^2+m_1^2} \, ,
\\ q_{2,0}^{(+)} &=& \sqrt{\vec{\ell}_2^2+m_2^2} \, ,
\\ q_{3,0}^{(+)} &=& \sqrt{\vec{\ell}_3^2+m_3^2} \, ,
\\ q_{4,0}^{(+)} &=& \sqrt{(\vec{\ell}_1+\vec{\ell}_2+\vec{\ell}_3)^2+m_4^2} \, ,
\\ q_{5,0}^{(+)} &=& \sqrt{(\vec{\ell}_1+\vec{\ell}_2)^2+m_5^2} \, .
\eeqn

In order to build the counter-term, we start looking at the single UV limit. As we explained before, the expansion in $\lambda$ has to be done up to order $3n-m$ around infinity, with $m$ number of internal lines depending on the $n$ divergent loop three-momenta $\vec{\ell}_\gamma$. Hence, we set $n=1$ for the single UV limit. For the momenta $\vec{\ell}_1$ and $\vec{\ell}_2$, we have $m=3$ and the amplitude is already integrable in the limits $|\vec{\ell}_1| \to \infty$ and $|\vec{\ell}_2| \to \infty$. However, for $\vec{\ell}_3$, it is $m=2$ so we need to expand the reduced amplitude up to order 1. Then, the corresponding local counter-term for the reduced amplitude is given by
\beqn
\nn {\cal A}_{RED, \uv, 3}^{(3)}&=&\frac{1}{q_{3,0,\uv}^{(+)}\left(q_{1,0}^{(+)}+q_{2,0}^{(+)}+q_{5,0}^{(+)}\right)} 
\\ &=&  \frac{1}{q_{3,0,\uv}^{(+)}\, \lambda_2} \, ,
    \label{eq:A3REDUV3_NMLT}
\eeqn
which leaves unchanged the causal threshold associated to $\lambda_2$.

Regarding the double UV limit ($n=2$), there are three cases. All of them have $m=4$, i.e. four propagators depend on $\{\ell_i,\ell_j\}$ for any pair $i,j$. Then, the expansion is carried out up to order $3n-m=2$ in $\lambda$. After applying the replacement rule $\mathcal{S'}_{\uv^\rho,\gamma}$ (with $\rho=2$ and $\gamma=\{\{1,2\},\{1,3\},\{2,3\}\}$) defined in Eq. (\ref{eq:Rep3GENERAL}) to 
\beq
({\cal A}^{(3)}_{RED})^{\prime}={\cal A}^{(3)}_{RED}-{\cal A}^{(3)}_{RED,\uv,3} \, ,
\eeq
and expanding up to order 2 in $\lambda$, we obtain
\beqn
\nn \mathcal{A}_{RED,\uv,12}^{(3)} &=&\frac{1}{Q_{2,\uv}} \left(\frac{1}{q_{12,0,\uv}^{(+)}}-\frac{1}{q_{3,0,\uv}^{(+)}} \right)
\\ &+& \frac{1}{Q_{2,\uv}^2}, 
\\ \nn \mathcal{A}_{RED,\uv,13}^{(3)} &=&\frac{1}{q_{1,0,\uv}^{(+)}} \left(\frac{1}{\tilde{Q}_{13,\uv}}-\frac{1}{2 q_{3,0,\uv}^{(+)}}\right)
\\ &+& \frac{1}{\tilde{Q}_{13,\uv}^2} \, ,
\\ \nn \mathcal{A}_{RED,\uv,23}^{(3)} &=&\frac{1}{q_{2,0,\uv}^{(+)}} \left(\frac{1}{\tilde{Q}_{23,\uv}}-\frac{1}{2 q_{3,0,\uv}^{(+)}}\right)
\\ &+& \frac{1}{\tilde{Q}_{23,\uv}^2} \, ,
\eeqn
where we introduced the short-hand notation
\beq
\tilde{Q}_{ij,\uv} = q_{i,0,\uv}^{(+)} + q_{j,0,\uv}^{(+)} +q_{ij,0,\uv}^{(+)} \, ,
\eeq
and $Q_{2\uv}$ corresponds to the UV-version of the causal threshold $\lambda_2$, i.e.
\beq
Q_{2,\uv} = q_{1,0,\uv}^{(+)} + q_{2,0,\uv}^{(+)} + q_{12,0,\uv}^{(+)} \, .
\eeq
Notice that $\tilde{Q}_{ij,\uv}$ is not directly related to a causal threshold of the original Feynman diagram.

Finally, in the triple UV limit, we start from the reduced amplitude without the simple and double UV counter-terms, i.e. 
\beqn
\nn ({\cal A}^{(3)}_{RED})^{\prime\prime}&=&({\cal A}^{(3)}_{RED})^{\prime}-{\cal A}^{(3)}_{RED,\uv,12}
\\ &-& {\cal A}^{(3)}_{RED,\uv,13} - {\cal A}^{(3)}_{RED,\uv,23} \, ,
\eeqn
and perform the expansion in $\lambda$ up to order 4. This is because $n=3$ (three simultaneous momenta are going to infinity) and $m=5$ (there are five internal lines). The resulting counter-term is lengthy, thus we present it in an ancillary file \cite{ZENODO}.

%%%%%%%%%%%%%%%%%%%%%%%%%%%%%%
\begin{figure}[!h]
\centering
\includegraphics[scale=0.95]{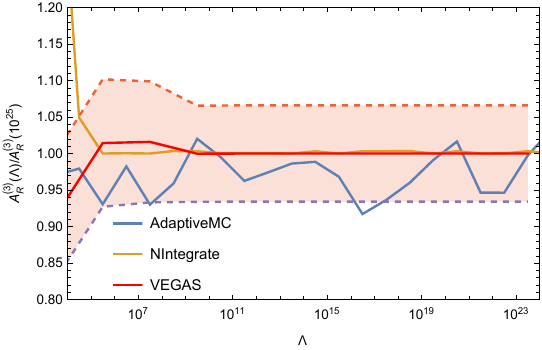}
\caption{Analysis of the numerical convergence for the 3-loop NMLT diagram with different masses, as a function of the cut-off $\Lambda$. The results are normalized to ${\cal A}^{(3)}_{R}(\Lambda=10^{25})$, using $m_1=4/10$, $m_2=m_3=m_4=1/10$, $m_5=4/10$, $\mu_\uv=1$ and neglecting units. The orange error band is obtained using \texttt{Vegas}.}
\label{fig:NMLT}
\end{figure}
%%%%%%%%%%%%%%%%%%%%%%%%%%%%%%

As in the other examples, we tested the convergence. In first place, we verify that the renormalized amplitude is integrable in the whole UV region, since it behaves as $1/|\vec{\ell}|^4$, $1/|\vec{\ell}|^7$ and $1/|\vec{\ell}|^{11}$ in the single, double and triple UV limits, respectively. Furthermore, we check that it behaves as $1/|\vec{\ell}|^5$ when only $|\vec{\ell}_1| \to \infty$ $|\vec{\ell}_2| \to \infty$, pointing towards a faster convergence.

%%%%%%%%%%%%%%%%%%%%%%%%%%%%%%
\begin{figure}[htb]
\centering
\includegraphics[scale=0.95]{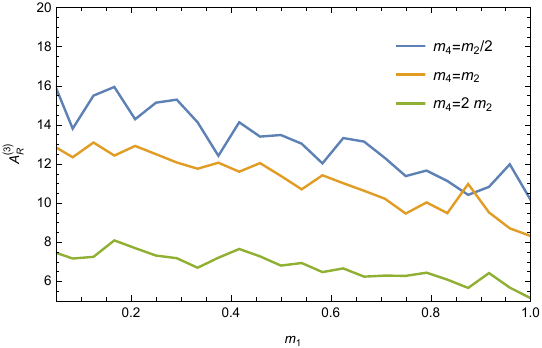}
\caption{Analysis of the mass dependence of the renormalized amplitude ${\cal A}^{(3)}_{R}$, with a fixed cut-off $\Lambda=10^{25}$. We set $m_2=m_3=4/10$, $m_5=3/10$ and $\mu_\uv=1$. Then, we vary $m_1 \in (1/48,1)$ and consider three different values of $m_4$: $m_4=m_2/2$ (blue), $m_4=m_2$ (orange) and $m_4=2m_2$ (green).}
\label{fig:NMLTmasas}
\end{figure}
%%%%%%%%%%%%%%%%%%%%%%%%%%%%%%

After that, we analyze the convergence varying the UV cut-off. In Fig. \ref{fig:NMLT}, we set $m_1=4/10$, $m_2=m_3=m_4=1/10$, $m_5=4/10$ and $\mu_\uv=1$, using three different methods: \texttt{NIntegrate} with \texttt{AdaptiveMonteCarlo} (blue line), the default \texttt{NIntegrate} (orange line) and \texttt{Vegas} (red line). The renormalized amplitude quickly reaches the asymptotic value, although oscillations are present (as in the 3-loop MLT case). By averaging over these three methods, the renormalized amplitude is ${\cal A}_R^{(3)}(\Lambda=10^{25})=35 \pm 13$, which leads to a relative error of ${\cal O}(70 \, \%)$. Still, the individual methods have large errors: ${\cal O}(100 \, \%)$ for the default \texttt{NIntegrate} scenario and ${\cal O}(20 \, \%)$ for \texttt{Vegas} (orange band). The estimation for \texttt{NIntegrate} with \texttt{AdaptiveMonteCarlo} is ${\cal O}(1 \, \%)$, using 5000000 points, and the method converges much faster than the other two strategies: for this reason, we set it as default for the next studies of this section. 

%%%%%%%%%%%%%%%%%%%%%%%%%%%%%%
\begin{figure}[htb]
\centering
\includegraphics[scale=0.95]{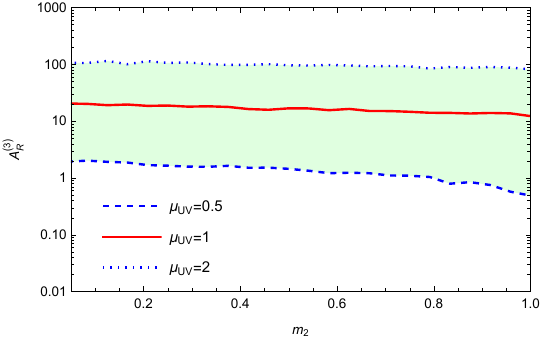}
\caption{Analysis of the mass and renormalization scale dependence $\mu_\uv$ of ${\cal A}^{(3)}_{R}$, with a fixed cut-off $\Lambda=10^{25}$ and $m_2 \in (1/48,1)$. We fix $m_1=2/10$, $m_3=m_4=3/10$ and $m_5=1/10$.}
\label{fig:NMLTrenormalization}
\end{figure}
%%%%%%%%%%%%%%%%%%%%%%%%%%%%%%

The next check consisted in studying the stability of the integral when varying the values of the masses. In Fig. \ref{fig:NMLTmasas}, we keep $m_2=m_3=4/10$, $m_5=3/10$ and $\mu_\uv=1$ fixed. We consider $m_1 \in (1/48,1)$, and three different scenarios: $m_4=m_2/2$ (blue), $m_4=m_2$ (orange) and $m_4=2m_2$ (green). Both the dependence in $m_1$ and $m_4$ are smooth, and the value of the integral decreases when heavier particles are considered. Again, this confirms the overall good quality of the local UV cancellation.

Finally, we examine the dependence on the renormalization scale $\mu_\uv$. We keep fixed $m_1=2/10$, $m_3=m_4=3/10$ and $m_5=1/10$, and the cut-off $\Lambda = 10^{25}$. Then, we vary $m_2$ within the range $(1/48,1)$, as we show in Fig. \ref{fig:MLT3renormalization}. We consider three different values of $\mu_\uv$: $1/2$ (dashed blue), $1$ (red) and $2$ (dotted blue). The band formed around the central value (i.e. for $\mu_\uv=1$) is very wide, covering roughly two orders of magnitude. Again, this indicates that the original integral is very UV-divergent and that the local counter-terms found with our procedure successfully neutralize these non-integrable terms.

%20240220: FINISHED!

%%%%%%%%%%%%%%%%%%%%%%%%%%%%%%%%%%%%%%%%%%%%%%%%%%%%%%%%%%%%%%%%%%%%%%%%%%%%%%%%%%%%%%
\section{Connection to BPHZ approach}
\label{sec:BPHZ}
As stated in the Introduction, techniques for getting rid of UV singularities are well-established in the literature. One relevant example in the context of this work is the BPHZ (Bogoliubov-Parasiuk-Hepp-Zimmermann) formalism. It was originally developed by Bogoliubov and Parasiuk in Ref. \cite{Bogoliubov:1957gp}, and consisted on the definition of an operator acting on diagrams with the purpose of locally removing the UV singularities. This operator, known as Bogoliubov's $R$-operator, relies on Taylor-like expansions in momentum space combined with graph theory techniques to identify overlapping singularities. It was shown that this strategy can successfully remove UV singularities for any renormalizable QFT \cite{Zimmermann:1969jj}, even at higher perturbative orders. Recently, there were efforts to define an extended operator, the $R^*$-operator \cite{Herzog:2017bjx}, capable of removing both IR and UV singularities.  

The purpose of this Section is comparing the BPHZ approach w.r.t. the local renormalization program within Causal Loop-Tree Duality. As a first step, let us consider a graph $\Gamma$. Then, the BPHZ $R$-operator acting on $\Gamma$ is given by 
\begin{equation}
R_{\gamma} \equiv 1-t_p^{\delta(\Gamma)} \, .
\end{equation}
The operator $t_p^{\delta(\Gamma)}$ symbolizes the Taylor expansion up to order $\delta(\Gamma)$ (the UV degree of divergence of the graph or subgraph), depending on the external momenta $p$, that extracts the divergent part of the diagram. It is important to notice that this Taylor expansion is equivalent to the expansion around the UV-propagator described in Sec. \ref{ssec:UVFelix} (as shown in App. \ref{app:A}): the operator $t_p$ is analogous to $L_\lambda$ in our formalism, choosing $\mu_{\uv}=m$.

Also, we observe that taking the UV limit of a set of loop-momenta $\delta$ corresponds to the application of the $R$-operator to a certain sub-diagram whose internal lines only depend on the loop-momenta $\ell_i \in \delta$. In particular, the $R$-operator applied on the whole graph $\Gamma$ is equivalent to the simultaneous UV limit of all loop-momenta. Thus, we appreciate a parallelism with the operators ${\cal S}'_{\uv,i}$ introduced in Sec. \ref{ssec:CausalLTD} to take the simultaneous UV limit of $i$ loop-momenta.

Besides, to discuss the removal of divergences of each subgraph $\gamma \in \Gamma$, let us write the decomposition 
\begin{equation}
I_{\Gamma}(p, k)=I_{\Gamma / \gamma} \, I_\gamma\left(p^\gamma, k^\gamma\right) \, ,
\end{equation}
where contracting $\gamma$ to a point within $\Gamma$ leads to the \emph{reduced diagram} $\Gamma / \gamma$. The external momenta $k^\gamma$ and $p^\gamma$ have to be chosen consistently with the parametrization of $\Gamma$ and the energy-momentum conservation at the vertices. Then, the UV expansion operator $t_\gamma$ in such a way that when it is applied to the amplitude $I_\Gamma$, it removes the sub-divergence coming from $\gamma$.

Once we identified common aspects of both strategies, we proceed to look into more detail the different operations performed. On one hand, the first step of our UV-expansion algorithm consists in removing the single UV limits. Using the notation introduced for BPHZ, the single UV counter-term can be re-written as
\beq
{\cal A}^{(L)}_{RED,1} = (1-\sum\limits_{i=1} t_{\gamma_i}) {\cal A}^{(L)}_{RED,0} \, .
\eeq
Then, the counter-terms for the double UV-limit take the form, 
\beq
{\cal A}^{(L)}_{RED,2} = (1-\sum\limits_{k=1}^{L-1} \sum\limits_{j>k}^{L} t_{\gamma_{jk}})(1-\sum\limits_{i=1}^L t_{\gamma_i}) {\cal A}^{(L)}_{RED,0} \, .
\eeq
Finally, iterating the procedure, the locally renormalized amplitude is given by
\beq
{\cal A}^{(L)}_{RED,L} = \prod_{l=1}^n(1-\sum\limits_{i=\{i_1,\ldots,i_n\}} t_{\gamma_{\{i_1,\ldots,i_n\}}}) {\cal A}^{(L)}_{RED,0} \, ,
\label{GeneralFinalAmp}
\eeq
where the summation is carried out over all ordered pairs $\{i_1,\ldots,i_n\}$.

On the other hand, let us consider Zimmerman's forest formula \cite{Zimmermann:1969jj}, i.e.
\begin{equation}
R_{\Gamma} I_{\Gamma}=\left(1-t_{\Gamma}\right) \sum_\alpha\left(\prod_{\gamma \in \mathcal{F}_\alpha}\left(-t_\gamma\right)\right) I_{\Gamma} \, .
\label{ZimmermanFF}
\end{equation}
Now, let us consider a graph without disjoint subgraphs, meaning that a couple of subgraphs can just be non-overlapping if one of the subgraphs is contained in the other. It was shown by Bergere and Zuber \cite{Bergere:1974zh} that the product of $t_\gamma$ operators of overlapping graphs applied on an amplitude is zero. So, it follows that the Zimmerman's forest formula Eq. (\ref{ZimmermanFF}) has the same form of Eq. (\ref{GeneralFinalAmp}). In the case that the diagram considered has disjoint subgraphs, the algorithm presented in this paper would need to be modified to recover Zimmerman's forest formula. In fact, in Eq. (\ref{GeneralFinalAmp}), there are some missing terms where the divergences of disjoint subgraphs are simultaneously removed, unlike in Eq. (\ref{ZimmermanFF}).

To end this comparison between both strategies, it should be noted that all the calculations and reasoning in this section are equally valid when working in three-dimensional Euclidean space. Therefore, the local renormalization proposed in this article by means of UV expansions within Causal Loop-Tree Duality is equivalent to the BPHZ approach, in the case of Feynman diagrams without disjoint subgraphs. Still, since our local renormalization techniques starts from the causal LTD representation, it has the advantage that the structure of the denominators is independent of the explicit momenta labelling: only on-shell energies appear (which, of course, implicitly depends on the momenta configuration).

%20240430: OK!

%%%%%%%%%%%%%%%%%%%%%%%%%%%%%%%%%%%%%%%%%%%%%%%%%%%%%%%%%%%%%%%%%%%%%%%%%%%%%%%%%%%%%%
\section{Conclusions and outlook}
\label{sec:Conclusions}
In this work, we explored techniques for achieving an \emph{integrand-level} renormalization of multiloop multileg scattering amplitudes. We started by reviewing a method that exploits the expansion of the integrand around the UV-propagator in the Minkowski space. Even if this technology was successfully tested in Refs. \cite{Sborlini:2016gbr,Sborlini:2016hat,Driencourt-Mangin:2019aix,Driencourt-Mangin:2019yhu} up to two-loops, going beyond this level posses additional difficulties. These difficulties are mainly related to the presence of new overlapped singularities introduced by counter-terms in the different UV limits. In this regard, we also showed that our method shares several nice properties with BPHZ renormalization program, in particular when dealing with graphs without disjoint subgraphs.

Thus, we exploited the nice properties of Loop Tree-Duality \cite{Catani:2008xa} to perform the UV-expansion in an Euclidean space. We took advantage of the so-called causal dual representations \cite{Verdugo:2020kzh} to expand around the infinite three-momentum regions inside the positive on-shell energies, which leads to more compact expressions. We tested the methodology with scalar Feynman integrals belonging to the Maximal Loop Topology (MLT) family at two and three loops, and the Next-to-Maximal Loop Topology (NMLT) at three loops. In all the cases, the counter-terms found locally cancel the non-integrable terms of the original amplitude in all the UV limits, hence rendering the expressions integrable in four space-time dimensions and without the need of introducing any additional regularization method.

One key aspect of the formalism presented in this work is the definition of \emph{reduced amplitudes}, and the subsequent application of the UV-expansion on them, instead of acting on the whole amplitude. In this way, the local renormalization does not alter the integration measure that appears in the causal dual representation; any multiloop multileg renormalized scattering amplitude reads
\beq
{\cal A}_{R,N}^{(L)} = \int_{\vec{\ell_1} \ldots \vec{\ell}_L} \, \sum_k \, \frac{1}{x_{L+k}} \left({\cal A}^{(L,k)}_{RED}-{\cal A}^{(L,k)}_{RED,\uv} \right) \, ,
\eeq
where we are summing over different topological families of order $k$. This will be particularly relevant in the context of the Four Dimensional Unsubtraction (FDU) \cite{Hernandez-Pinto:2015ysa,Sborlini:2016gbr,Sborlini:2016hat} approach, since the dual (i.e. virtual after the application of LTD) contribution is combined, at integrand-level, with the real radiation. This real-dual combination involves a kinematical mapping, and keeping a phase-space measure within the contributions coming from loops plays a crucial role in the local cancellation of IR divergences. Thus, in sight of a unified framework to compute physical observables at higher-order directly in four space-time dimensions \cite{Ramirez-Uribe:2024rjg,LTD:2024yrb}, our findings regarding local renormalization from causal LTD might be very helpful. 

%20240220: FINISHED!!

%%%%%%%%%%%%%%%%%%%%%%%%%%%%%%%%%%%%%%%%%%%%%%%%%%%%%%%%%%%%%%%%%%%%%%%%%%%%%%%%%
\section*{Acknowledgements}
This article is dedicated to the memory of Stefano Catani, one of the most prominent contemporary scientists in the field of theoretical high-energy physics and the original developer of the Loop-Tree Duality formalism. We would like to thank Germ\'an Rodrigo, Roger Hern\'andez-Pinto and Leandro Cieri for the fruitful discussion about Loop-Tree Duality and for providing us very useful comments about the manuscript. This work was supported by the Spanish Government (Agencia Estatal de Investigaci\'on MCIN/AEI/ 10.13039/501100011033) Grants No. PID2019-105439GB-C22, PID2020-114473GB-I00, PID2022-141910NB-I00 and Generalitat Valenciana Grants No. PROMETEO/2021/071 and ASFAE2022/009. GS is partially supported by EU Horizon 2020 research and innovation programme STRONG-2020 project under Grant Agreement No. 824093 and H2020-MSCA-COFUND USAL4EXCELLENCE-PROOPI-391 project under Grant Agreement No 101034371.

%%%%%%%%%%%%%%%%%%%%%%%%%%%%%%%%%%%%%%%%%%%%%%%%%%%%%%%%%%%%%%%%%%%%%%%%%%%%%%%%%%%%%%
\appendix
\section{Equivalence of momentum expansions in the UV limit}
\label{app:A}
In order to show the equivalence between the expansion around the UV-propagator (as described in Sec. \ref{ssec:UVFelix}) and BPHZ renormalization in Minkowski spacetime, let us consider a propagator depending on an arbitrary number of loop and external momenta. 
Let us define the propagator of the $j$-th internal line as
\beq
    \Delta (k_j,m_j)=\frac{1}{k_j^2-m_j^2} \, ,
    \label{defProp}
\eeq
where $k_j=\sum\limits_{n \in \delta_j} \ell_n + \sum\limits_i p_i$. $\delta_j$ is the set of indices of loop momenta $\ell$ on which the $j$-th internal four-momentum depends. The summation over $i$ also runs through all external momenta on which it depends.
Let's consider the UV limit of some loop momenta, whose indices define the set $\gamma$.
Applying the replacements given in Eqs. (\ref{eq:ReemplazoSIMPLE})-(\ref{eq:ReemplazoDOBLE}) (depending whether it is a simple or multiple UV limit) and setting $\mu_{\uv}=m_j$, the propagator in Eq. (\ref{defProp}) becomes 
\beqn
    \nn & \Delta (k_j,m_j)|_{{\cal S}_{UV,\gamma}}=\frac{1}{\left(\lambda \sum\limits_{\substack{n \in \delta_j \\ n \in \gamma}} \ell_n + \sum\limits_{\substack{m \in \delta_j \\  m \notin \gamma}} \ell_m + \sum\limits_i p_i\right)^2 -\lambda^2m_j^2}
    \\ &= \frac{x^2}{\left(\sum\limits_{\substack{n \in \delta_j \\ n \in \gamma}} \ell_n + x\left( \sum\limits_{\substack{m \in \delta_j \\  m \notin \gamma}} \ell_m + \sum\limits_i p_i \right)\right)^2 -m_j^2} \, ,
    \label{RepPropBPHZ}
\eeqn
where the replacement $x=\frac{1}{\lambda}$ is carried out. This way, a Taylor expansion at $\lambda \rightarrow \infty$ corresponds to expanding $x$ around 0. After this expansion, the final result is recovered by taking the limit $x \rightarrow 1$.

On the other hand, it can be shown by a change of variables that
\beqn
    x^n \frac{d^n}{dx^n}=\left(\sum\limits_{\substack{m \in \delta_j \\  m \notin \gamma}} l_m \frac{\partial}{\partial l_m} + \sum\limits_i p_i \frac{\partial}{\partial p_i}\right)^n \, .
\eeqn
Then, the Taylor expansion of the propagator up to order $N$ can be obtained from
\beqn
    \sum\limits_{n=0}^N \frac{1}{n!} \left(\sum\limits_{\substack{m \in \delta_j \\  m \notin \gamma}} l_m \frac{\partial}{\partial l_m} + \sum\limits_i p_i \frac{\partial}{\partial p_i}\right)^n \, \Delta(k_j,m_j) \, .
\eeqn
Let us now define $\bar{p_i}$, which can be an external momentum or a loop momentum that doesn't belong to $\gamma$. In other words, the momenta labelled by $\bar{p_i}$ are kept finite.
Then, we can write
\beqn
    \nn & \left(\sum\limits_{\substack{m \in \delta_j \\  m \notin \gamma}} l_m \frac{\partial}{\partial l_m} + \sum\limits_i p_i \frac{\partial}{\partial p_i}\right)^n \Delta (k_j,m_j)
    \\ &= \left(\sum\limits_i \bar{p_i} \frac{\partial}{\partial \bar{p_i}} - \sum\limits_{s \notin \delta_j} l_s \frac{\partial}{\partial l_m} \right)^n \Delta (k_j,m_j)
    \\ \nn &= \left(\sum\limits_i \bar{p_i} \frac{\partial}{\partial \bar{p_i}}\right)^n \Delta (k_j,m_j) =  \bar{p}_{i_1}^{\mu_1} \cdots \bar{p}_{i_l}^{\mu_l} \frac{\partial^l \Delta (k_j,m_j)}{\partial \bar{p}_{i_1}^{\mu_1} \cdots \partial \bar{p}_{i_l}^{\mu_l}} .
\eeqn
Therefore, it is shown that the UV expansion algorithm proposed in Sec. \ref{sec:LocalReno} taking $\mu_{\uv}=m$ and applied on a single propagator corresponds exactly to the Taylor expansion performed within the Bogoliubov's $R$-operation in the BPHZ formalism. 
The argument presented here can be straightforwardly generalized to deal with an amplitudes with an arbitrary number of propagators.

%20240427: Revisado, OKK!!

%%%%%%%%%%%%%%%%%%%%%%%%%%%%%%%%%%%%%%%%%%%%%%%%%%%%%%%%%%%%%%%%%%%%%%%%%%%%%%%%%%%%%%

%\bibliography{refs}

%merlin.mbs apsrev4-1.bst 2010-07-25 4.21a (PWD, AO, DPC) hacked
%Control: key (0)
%Control: author (8) initials jnrlst
%Control: editor formatted (1) identically to author
%Control: production of article title (-1) disabled
%Control: page (0) single
%Control: year (1) truncated
%Control: production of eprint (0) enabled
%

\end{document}